\documentclass[12pt]{article}
\usepackage{amsmath}

\usepackage{tikz}
\usetikzlibrary{shapes}

\usepackage{array}
\usepackage{amsthm}
\usepackage{amssymb}

\usepackage{times}
\usepackage{graphicx}
\usepackage{color}
\usepackage{multirow}
\usepackage[authoryear,sort]{natbib}
\usepackage{rotating}
\usepackage{bbm}
\usepackage{latexsym}

\newcommand{\captionfonts}{\normalsize}

\makeatletter  
\long\def\@makecaption#1#2{%
  \vskip\abovecaptionskip
  \sbox\@tempboxa{{\captionfonts #1: #2}}%
  \ifdim \wd\@tempboxa >\hsize
    {\captionfonts #1: #2\par}
  \else
    \hbox to\hsize{\hfil\box\@tempboxa\hfil}%
  \fi
  \vskip\belowcaptionskip}
\makeatother   

\usepackage[colorlinks=true,allcolors=blue]{hyperref}

\newcommand\circled[1]{%
	\tikz[baseline=(X.base)] 
	\node (X) [draw, shape=circle, inner sep=0pt] {#1};}

\newcommand\squared[1]{%
	\tikz[baseline=(X.base)] 
	\node (X) [draw, shape=rectangle, inner sep=1.1pt] {#1};}

\newcommand\acircled[1]{%
	\tikz[baseline=(X.base)] 
	\node (X) [draw, shape=circle, inner sep=-.7pt] {#1};}

\newcommand\asquared[1]{%
	\tikz[baseline=(X.base)] 
	\node (X) [draw, shape=rectangle, inner sep=0.4pt] {#1};}

\newcommand\atriangled[1]{%
	\tikz[baseline=(X.base)] 
	\node (X) [draw,regular polygon, regular polygon sides=3,inner sep=0.1pt] {#1};}

\newcommand{\SCA}{\makebox[2.3ex][l]{\circled{\makebox[1.6ex][c]{A}}}}
\newcommand{\SCB}{\makebox[2.3ex][l]{\circled{\makebox[1.6ex][r]{B}}}}
\newcommand{\SBA}{\makebox[2.3ex][l]{\squared{\makebox[1,6ex][c]{A}}}}
\newcommand{\SBB}{\makebox[2.3ex][l]{\squared{\makebox[1.6ex][r]{B}}}}
\newcommand{\SA}{\mathrm{A}}
\newcommand{\SB}{\mathrm{B}}
\newcommand{\SCirc}{\makebox[2.2ex][c]{\acircled{\makebox[2ex][c]{\phantom{\rule[0ex]{0ex}{1.6ex}}}}}}
\newcommand{\SBox}{\makebox[2.2ex][c]{\asquared{\makebox[1.6ex][c]{\phantom{\rule[0ex]{0ex}{1.6ex}}}}}}
\newcommand{\STri}{\makebox[2.2ex][c]{\atriangled{\makebox[.8ex][c]{\phantom{\rule[0ex]{0ex}{.8ex}}}}}}

\makeatletter
\DeclareRobustCommand{\text}{%
  \ifmmode\expandafter\text@\else\expandafter\mbox\fi}
\let\nfss@text\text
\def\text@#1{{\mathchoice
  {\textdef@\displaystyle\f@size{#1}}%
  {\textdef@\textstyle\f@size{#1}}%
  {\textdef@\textstyle\sf@size{#1}}%
  {\textdef@\textstyle \ssf@size{#1}}%
  \check@mathfonts
  }%
}
\def\textdef@#1#2#3{\hbox{{%
                    \everymath{#1}%
                    \let\f@size#2\selectfont
                    #3}}}
\makeatother

\mathchardef\myminus="2D

\newcommand{\mysp}{\hspace{-0.05em}}
\makeatletter
\newcommand{\rs}[1]{%
  \@tfor\next:=#1\do{\next\mysp}%
}
\newcommand{\rl}[1]{%
  \@tfor\next:=#1\do{\next\!}%
}
\makeatother

\newcommand{\dini}{\Delta I}
\newcommand{\dinit}{\delta I}

\newcommand{\dinid}{\dini^{D\!}}

\newcommand{\dinidl}{\dini^{D\!L\!}}

\newcommand{\dinidd}{\dini^{B\!}}
\newcommand{\dinitdl}{\dinit}
\newcommand{\dinidlt}[1]{\dinitdl( #1 \theta)}
\newcommand{\dinidltsub}[2]{\dinitdl_{#2}( #1 \theta)}
\newcommand{\dint}{\nabla I}
\newcommand{\dintdl}{\dint^{D\!L\!}}

\newcommand{\jth}{j\textth}
\newcommand{\nth}{n\textth}
\newcommand{\kth}{k\textth}

\newcommand{\VR}{\mathbf{R}}

\newcommand{\VS}{\mathbf{S}}
\newcommand{\SSE}{S}
\newcommand{\VSD}{\tilde{\VS}}

\newcommand{\VSNI}{\VSD}

\newcommand{\MSEnc}[1][\phantom{1}]{\MSEncnj{N}{#1}}
\newcommand{\MSEncnj}[2]{\VS^{\mathtt{[#1]}}_{#2}}
\newcommand{\MSDec}[1][\phantom{1}]{\MSDecnj{N}{#1}}
\newcommand{\MSDecnj}[2]{\VSD^{\mathtt{[#1]}}_{#2}}

\newcommand{\MR}[1][\phantom{1}]{\MRnj{N}{#1}}
\newcommand{\MRnj}[2]{\VR^{\mathtt{[#1]}}_{#2}}

\newcommand{\PSNIerr}{P_e^{\mathtt{[N]}}}
\newcommand{\PSNIerrj}{P_{e,j}^{\mathtt{[N]}}}

\newcommand{\SJI}{\mathbf{S}^{J\!I}}
\newcommand{\SPI}{\mathbf{S}^{P\!I}}

\newcommand{\textth}{\textsuperscript{th}}

\newcommand{\comp}[1]{\bar{#1}}

\newcommand{\alphaj}{\alpha_j}
\newcommand{\alphac}{\comp{\alpha}}
\newcommand{\alphacj}{\comp{\alpha}_j}

\newcommand{\qc}{\comp{q}}
\newcommand{\qcac}{\qc\,\alphac}

\newcommand{\qa}{q\,\alpha}

\newcommand{\qj}{q_j}
\newcommand{\qcj}{\qc_j}

\newcommand{\kqj}{\frac{\qcj}{\qj}}

\newcommand{\kalphaj}{\frac{\alphacj}{\alphaj}}

\newcommand{\kalphajt}{\lrpar{\kalphaj}^\theta}

\newcommand{\Hopb}{\mathcal{H}}

\newcommand{\rdsarg}{(\VR|\VS)}

\newcommand{\sdrarg}{(\VS|\VR)}

\newcommand{\mtp}{\mysp\texttt{+}\mysp}
\newcommand{\sj}{{1,\ldots,J}}
\newcommand{\sjsc}{{1;\ldots;J}}
\newcommand{\spj}{{1 \mtp \ldots \mtp J}}

\newcommand{\infoj}{I_\sj}
\newcommand{\dintdlj}{\dintdl_\sjsc}
\newcommand{\dinidlj}{\dinidl_\sj}

\newcommand{\dinidlpj}{\dinidl_\spj}

\newcommand{\spinf}{{1\mtp\ldots\mtp \infty}}
\newcommand{\sdinf}{{1,\ldots,\infty}}
\newcommand{\sintinf}{{1;\ldots;\infty}}

\newcommand{\sdrtharg}{(\VS|\VR,\theta)}

\newlength{\mytextsize}

\newcommand{\llnapprox}{\underset{\scriptscriptstyle J\gg 1}{\approx}}

\newcommand{\defeq}{\overset{\scriptscriptstyle \mathrm{def}}{=}}

\newcommand{\PNI}{P^{N\!I}}
\newcommand{\PNIs}{P^{\rl{NI}}}

\newcommand{\lrpar}[1]{\left(#1\right)}
\newcommand{\lrcor}[1]{\left[#1\right]}

\newcommand{\lrang}[1]{\left\langle#1\right\rangle}

\newcommand{\PrS}{P(\VS)}

\newcommand{\PSdR}{P\sdrarg}

\newcommand{\info}[2]{I{(#1;#2)}}

\newcommand{\eref}[1]{\hyperref[#1]{(\ref{#1})}}
\newcommand{\fref}[1]{\hyperref[#1]{Fig.~\ref{#1}}}
\newcommand{\sref}[1]{\hyperref[#1]{Section~\ref{#1}}}
\newcommand{\tref}[1]{\hyperref[#1]{Table~\ref{#1}}}

\begin{document}

\begin{center}
{\LARGE Disambiguating the Role of Noise Correlations\\[1ex] When Decoding Neural Populations Together}\\[3ex]
\textbf{\large Hugo Gabriel Eyherabide}\\[3ex]
\end{center}

{
\linespread{1}\selectfont
\begin{center}
\textit{Department of Mathematics and Statistics, Department of Computer Science, and Helsinki Institute for Information Technology, University of Helsinki \linebreak Gustaf H\"allstr\"omin katu 2b, FI00560, Helsinki, Finland}\\[3ex]
\end{center}
}

\begin{center}
{\textbf{Email:} neuralinfo@eyherabidehg.com}\hfill {\textbf{Weppage:} eyherabidehg.com}
\end{center}


\begin{center} {\bf \large Abstract} \end{center}
One of the most controversial problems in neural decoding is quantifying the information loss caused by ignoring noise correlations during optimal brain computations. For more than a decade, the measure here called $ \dinidl $ has been believed exact. However, we have recently shown that it can exceed the information loss $ \dinidd $ caused by optimal decoders constructed ignoring noise correlations. Unfortunately, the different information notions underlying $ \dinidl $ and $ \dinidd $, and the putative rigorous information-theoretical derivation of $ \dinidl $, both render unclear whether those findings indicate either flaws in $ \dinidl $ or major departures from traditional relations between information and decoding. Here we resolve this paradox and prove that, under certain conditions, observing $ \dinidl{>}\dinidd $ implies that $ \dinidl $ is flawed. Motivated by this analysis, we test both measures using neural populations that transmit independent information. Our results show that $ \dinidl $ may deem noise correlations more important when decoding the populations together than when decoding them in parallel, whereas the opposite may occur for $ \dinidd $. We trace these phenomena back, for $ \dinidd $, to the choice of tie-breaking rules, and for $ \dinidl $, to unforeseen limitations within its information-theoretical foundations. Our study contributes with better estimates that potentially improve theoretical and experimental inferences currently drawn from $ \dinidl $ without noticing that it may constitute an upper bound. On the practical side, our results promote the design of optimal decoding algorithms and neuroprosthetics without recording noise correlations, thereby saving experimental and computational resources.

\pagebreak

\section{Introduction}\label{sec::intro}

Noise correlations modulate the coactivation of neurons and neural populations at multiple levels in the brain, potentially introducing information that cannot be decoded without knowing their strength and structure. To test this hypothesis, previous studies have assessed, for example, whether correlations are strong, vary across stimuli or experimental tasks, increase the amount of encoded information, or shape spike-triggered averages and decoding filters \citep{gawne1993,meister1995,warland1997,nirenberg1998,panzeri2001,abbott1999,brenner2000,schneidman2003,eyherabide2008,quiroga2009}. These findings were thought to imply that ignoring noise correlations during decoding must cause an information loss.

This conclusion was challenged by \citet{nirenberg2001} who proposed to test the hypothesis directly from the decoder, observer or organism perspective \citep{bialek1991,jaynes2003}. To that end, they derived an information-theoretical measure of correlation importance here called $ \dinid $ \citep[also called $ \Delta I $, $ I_{\rs{cordep}} $, $ \Delta I_1 $, and $ \Delta I_{N\!I}^D $][]{nirenberg2003,montani2007,ince2010,eyherabide2013}. However, the measure has been perceived as an upper bound, partially because it may potentially exceed the information loss $ \dinidd $ caused by optimal decoders constructed assuming noise independence and even the transmitted information \citep{schneidman2003,latham2005,oizumi2009,oizumi2010,eyherabide2013,latham2013}.

These putative limitations were seemingly solved by \citet{latham2005} who derived a new information-theoretical measure of correlation importance here called $ \dinidl $ \citep[also denoted $ \Delta I^* $ and $ \Delta I_{N\!I}^{D\!L} $ in neural decoding, and analogous to $ \Phi^*$ within integrated information theory of consciousness;][]{latham2005,eyherabide2013,oizumi2016}. Initially, this measure was employed to justify $ \dinid $ as an upper-bound of correlation importance \citep{latham2005}. However, its applications have been extended to included the study of medium-to-large populations, thereby overcoming the putative overestimation produced by $ \dinid $; higher-order neural correlations, by combining $ \dinidl $ with maximum-entropy methods; integrated-information theory of consciousness, introducing the decoder perspective and the first measure that seemingly fulfills the theoretical requirements within the field; and neural stochastic codes from the decoding perspective \citep{oizumi2009,ince2010,oizumi2010,latham2013,oizumi2016,eyherabide2016b}.

Since its introduction, $ \dinidl $ has been regarded as the exact information loss caused by ignoring noise correlations in optimal decoding \citep{latham2005,oizumi2009,ince2010,oizumi2010,latham2013,oizumi2016}. However, we have recently shown that, like its predecessor, $ \dinidl $ can also exceed $ \dinidd $ \citep{eyherabide2013}. Unfortunately, due to the rigorous information-theoretical derivation of $ \dinidl $ and the different information notions underlying $ \dinidl $ and $ \dinidd $, 
whether the aforementioned numerical comparison either reveals major departures from traditional views on the relation between information and decoding, or constitutes an indication that $ \dinidl $ is flawed, remains an open question.

To answer this question, we first disentangle the information notion underlying $ \dinidl $, here called communication information, from the one that we argue underlies $ \dinidd $, which we call axiomatic information. Taking their differences into account, we determine under which conditions observing that $ \dinidd{<}\dinidl  $ implies that $ \dinidl $ is flawed and overestimates the communication information loss. We also address whether this conclusion can be reached even if $ \dinidl $ does not exceed $ \dinidd $. To that end, we study neural populations that transmit independent information, and show that $ \dinidl $ grows when decoding them together, as opposed to decoding them in parallel. This paradoxical growth, which can reach about $ 100\,\% $ of transmitted information, is here shown to stem from unforeseen information-theoretical limitations in the derivation of $ \dinidl $. Surprisingly, we find exactly the opposite phenomena when using $ \dinidd $, and trace it back to the choice of tie-breaking rules employed during the decoder construction. Our study shows, for the first time, that none of these measures need be additive when information is independent, and most importantly, that $ \dinidl $ need not be exact and may overestimate the communication information loss. Above all, we contribute with tight estimates of communication information losses, thereby potentially improving the accuracy of previous and future theoretical and experimental inferences drawn from them. On the practical side, our results open up new possibilities for simplifying, with tolerable information losses, the computational models that underlie the design of brain-machine interfaces and neuroprosthetics, and for reducing the amount of resources required to study brain computations and information integration \citep{nirenberg2001,nirenberg2003,latham2005,quiroga2009,eyherabide2013,aflalo2015,zhang2016,bouton2016}.

\section{Materials and methods}

\subsection{Notation}\label{met::sec::notation}

Sensory stimuli are here characterized by vectors $ \VS{=}[S_1,\ldots,S_J]$ of $ J $ components, where each $ \jth $ component $ S_j $ represents the value adopted by a different feature. For example, most of our hypothetical experiments employ the following four stimuli: $ \SBA $, $ \SBB $, $ \SCA $ and $ \SCB $. These stimuli can be characterize using vectors $ \VS{=}[S_1,S_2] $ of two components (i.e. $ J{=}2 $), where the first component denotes the type of frame (i.e., $ S_1{\in}\{\SBox,\SCirc\}) $, and the second component, the type of letter (i.e., $ S_2{\in}\{\SA,\SB\} $).

Neural responses are here characterized by vectors $\VR{=}[\VR_1,\ldots,\VR_J] $ of $ J $ components. Each $ \jth $ component $ \VR_j$ typically characterizes those aspects of the neural responses (e.g., first-spike latency and spike counts either in the individual or concurrent activity of all neurons; phase and amplitudes in local field potentials or sensor signals from brain-imaging devices; etc.) that are sensitive to the value adopted by the $ \jth $ stimulus feature. To that end, each $ \VR_j$  is here characterized as a vector of $ K_j $ components, namely $ \VR_j{=}[R_j^1,\ldots,R_j^{K_j}] $, where each $ \kth $ component $ R_j^k $ denotes the value adopted by the $ \kth $ response aspect in the set of response aspects that are sensitive to the $ \jth $  stimulus feature $ S_j $. However, to improve readability, the subscript $ j $ will often be eliminated when $ J{=}1 $.

As an example, imagine an experiment with two populations of two neurons each, that fire in response to the stimuli mentioned above. The first population is only sensitive to frames and the second population is only sensitive to letters. In this experiment, we will denote the concurrent responses of all neurons as $ \VR{=}[\VR_1,\VR_2] $. Here, $ \VR_1 $ and $ \VR_2 $ denote the concurrent responses of all neurons in the first and the second population, respectively. In addition, $ \VR_1{=}[R_1^1,R_1^2] $, with $ R_1^1 $ and $ R_1^2 $ denoting the responses of the first and the second neuron in the first population; $ \VR_2 $ is defined analogously to $ \VR_1 $. Should the four neurons have been sensitive to all stimulus features, we would have denoted their concurrent responses as $ \VR{=}[R^1,R^2,R^3,R^4] $.

\subsection{Neural encoding}\label{met::sec::neuenc}

Transforming $ \VS $ into $ \VR $ is called encoding \citep{panzeri2010}. Because the same $ \VS $ may elicit different $ \VR $s, and the same $ \VR $ may occur for different $ \VS $s, both $ \VS $ and $ \VR $ are often treated as random variables with joint probabilities $ P(\VS,\VR) $. The Shannon or mutual information $ I(\VS;\VR)$ encoded in $ \VR $ about $ \VS $ is given by the following

\begin{equation}\label{met::eq::encinfo}
I(\VS;\VR) = \sum_{\VS,\VR}{P(\VS,\VR)\,\ln \frac{P(\VS|\VR)}{P(\VS) \,P(\VR)}}\, ,
\end{equation}

\noindent where $ \ln $ denotes natural logarithms. Consequently, absolute information values are measured in units of nats, as opposed to units of bits \citep[the conversion from nats to bits only requires to divide by $ \ln 2 $;][]{cover2006}. However, our choice of natural logarithms does not affect information ratios, and simplifies calculations and notation.

\subsection{Noise correlations}\label{met::sec::noisecorr}

The responses of $ K $ neurons are deemed noise independent (NI) when the following condition always holds

\begin{equation}\label{met::eq::noiseind}
P(\VR|\VS) = \prod_{k=1}^{K}{P(R^k|\VS)} \defeq \PNIs(\VR|\VS)\, . 
\end{equation}

\noindent Otherwise, the responses are deemed noise correlated. Here, $ \defeq $ denotes a definition; and $ \prod $, a product. This definition can be traced back to \citet{schneidman2003} and takes into account noise correlations at all orders \citep{latham2013}, as opposed to those based on linear or nonlinear correlations \citep{pereda2005,cohen2011}. However, contrary to previous studies \citep{nirenberg2003,latham2005,meytlis2012,delis2013}, here we note that this definition need not be equivalent to those that additionally average across all stimuli \citep{gawne1993,womelsdorf2012}, which potentially confuse noise correlations with activity correlations \citep{schneidman2003} and are prone to cancellation effects \citep{nirenberg2003}.

\subsection{Neural decoding}\label{met::sec::neudec}

Transforming $ \VR $ into estimated stimuli $ \VSD $ (or into perceptions, decisions and actions) is called decoding \citep{panzeri2010}. Analogous to $ \VS $ and $ \VR $, both $ \VS $ and $ \VSD $ are often treated as random variables with joint probabilities $ P(\VS,\VSD) $, also called confusion matrix \citep{quiroga2009,ince2010,rolls2011,delis2013}. In this study, we focus on optimal decoders, also known as Bayesian or maximum-a-posteriori decoders, ideal homunculus, ideal or Bayesian observers, and optimal-unbiased or maximum-likelihood discrimination \citep{bialek1987,knill1996,oram1998,ernst2002,simoncelli2009,geisler2011,zhang2016}. These decoders map each $ \VR $ into $ \VSD$ as follows

\begin{equation}\label{met::eq::bayesdec}
\VSD = \arg \max_{\VS}{P(\VS|\VR)}\, ,
\end{equation}

\noindent where $ P(\VS|\VR)$ is computed via Bayes' rule \citep{eyherabide2013}.

When neurons are noise correlated, decoding their concurrent responses $ \VR $ using the exact $ P(\VR|\VS) $ can become experimentally and computationally intractable even for the brain. For this reason, previous studies have proposed to construct optimal decoders assuming that neurons are NI, here called optimal NI decoders \citep{eyherabide2013}, but also known as weak-coupling or independent models, weak observers, and naive Bayes classifiers \citep{landy1995,knill1996,duda2000,nirenberg2001,nirenberg2003,quiroga2009,meytlis2012}. These decoders map each $ \VS $ into $ \VSNI $ as follows

\begin{equation}\label{met::eq::bayesnidec}
\VSNI = \arg \max_{\VS}{\PNIs(\VS|\VR)}\, ,
\end{equation}

\noindent with $ \PNIs(\VS|\VR)$ computed from $\PNIs(\VR|\VS) $ via Bayes' rule.

\subsection{Decoding perspective}\label{met::sec::perspective}

This study assesses the role of noise correlations from the decoding perspective. Within it, the importance of noise correlations is measured as the losses caused by decoding the actual neural responses assuming that neurons are NI. Noise correlations are deemed important if the losses are significant, and inessential if they are not \citep[see][and references therein]{eyherabide2013}. The decoding perspective if often confused with other approaches that measure correlation importance, for example, by comparing the information encoded in real responses with the one encoded in surrogate NI responses (responses generated assuming that neurons are NI). Even when using decoders \citep{nirenberg1998,quiroga2009,delis2013}, these approaches need be neither conceptually nor quantitatively related to the decoding perspective \citep{nirenberg2003,latham2005,averbeck2006}. Avoiding such confusion is fundamental to correctly interpreting our results and conclusions.

\subsection{Measures of correlation importance}\label{met::sec::measures}

Even within the decoding perspective, the choice of correlation-importance measure remains controversial. Here we will study the relation between the following three commonly-used measures

\begin{align}
\dinidd(S,\VR) & = \sum_{S,\VR}{P(S,\VR)\,\ln \frac{P(S|\VR)}{P(S|\VSNI)}} \label{met::def::dinidd}\\
\dinid(S,\VR) & = \sum_{S,\VR}{P(S,\VR)\,\ln \frac{\PSdR}{\PNIs\sdrarg}}\label{met::def::dinid}\\
\dinidl(S,\VR) & = \min_{\theta}{\sum_{S,\VR}{P(S,\VR)\,\ln \frac{\PSdR}{\PNIs\sdrtharg}}}\label{met::def::dinidl}\defeq \min_{\theta}{\dinidlt{\VS;\VR|}}\, ,
\end{align}

\noindent where $ \VSNI $ is the output of an optimal NI decoder (\sref{met::sec::neudec}), and 

\begin{equation}\label{met::eq::qsrtheta}
\PNIs\sdrtharg \propto \PrS\,\PNIs\rdsarg^{\theta} \, ,
\end{equation}

\noindent with the convention that $ 0^0{=}0 $ for automatically overcoming the drawbacks of previous definitions found in \citep{eyherabide2013}. These measures have been previously related to the information loss caused by ignoring noise correlations in optimal decoding \citep{nirenberg2003,latham2005,ince2010}. Accordingly, the decoded information when ignoring noise correlations in optimal decoding has previously been quantified by subtracting these measures from the encoded information \citep{oizumi2009,oizumi2010}. Further details about their derivations and interpretations are here postponed until \sref{sec::results}, where we will reassess them in the context of the present study.

\subsection{Independent information}\label{met::sec::indinfo}

The notion of independent information has previously been given different definitions that need not be interchangeable. In this study, we say that the responses $ \VR_1,\ldots,\VR_J $ of $ J $ neural populations transmit independent information when they fire independently and selectively to $ J $ independent stimulus features $ S_1,\ldots,S_J $ \citep{fano1961}. Using the notation introduced in \sref{met::sec::notation}, our definition implies the following

\begin{equation}\label{met::eq::indjoint}
P(\VS,\VR) = \textstyle\prod_{j=1}^J{P(S_j,\VR_j)} \, .
\end{equation} 

\noindent When \eref{met::eq::indjoint} holds, each pair $ [S_j,\VR_j] $ is here said to constitute an independent-information channel or stream.

Under our definition, the information carried by independent-information streams is additive \citep{fano1961}, namely 

\begin{equation}\label{met::eq::infoadd}
I(\VS,\VR)= \sum_{j=1}^{J}{I(S_j;\VR_j)}\, .
\end{equation}

\noindent For simplicity, hereinafter we abbreviate the notation for the arguments of any information measure $ X $ as follows

\begin{align}
\makebox[6ex][l]{$ X_{1,\ldots,J} $}&\defeq X(S_1,\ldots,S_J;\VR_1,\ldots,\VR_J)\\
\makebox[6ex][l]{$ X_j $} &\defeq X(S_j,\VR_j)\\
\makebox[6ex][l]{$ X_\spj $}&\defeq X_1+\cdots+X_J
\end{align}

\noindent Using these abbreviations, the additivity of the information carried by independent-information streams can be simply put as $ I_{\sj}{=}I_{\spj} $.

This property has often been used as the actual definition of independent information \citep{brenner2000,schneidman2003,schneidman2011,rolls2011}, whereas other definitions only require that populations are asymptotically or conditionally independent \citep{gawne1993,samengo2000,cover2006}. Our definition is more stringent than those and ensures that independent information remains independent after arbitrary parallel transformations of the form $ [\tilde{S}_j,\tilde{\VR}_j]=f_j(S_j,\VR_j)$, namely

\begin{equation}\label{se::eq::infoaddtransform}
P\bigl(\tilde{S}_1,\tilde{\VR}_1,\ldots,\tilde{S}_J,\tilde{\VR}_J\bigr) = \textstyle\prod_{j=1}^{J}{P\bigl(\tilde{S}_j,\tilde{\VR}_j\bigr)}\, .
\end{equation}

\noindent This property plays a fundamental role in our study, and it also holds for the definitions given in \citet{cover2006} and \citet{eyherabide2010FCN}. However, our definition is more general than those for the noise in each $ \jth $ stream is allowed to depend on the $ \jth $ feature.

\section{Results}\label{sec::results}

\subsection{Underlying information notions need not be reliably related}\label{res::sec::difnotion}

Quantifying the information loss caused by ignoring noise correlations from the decoding perspective remains controversial, oftentimes due to unfulfilled expectations about the relation between different measures of information loss. One of the most important unfulfilled expectations concerns the measures $ \dinidl $ and $ \dinidd $ (\sref{met::sec::measures}). Because $ \dinidl $ continues to be considered the exact information loss in optimal NI decoding (\sref{sec::intro} and references therein), it may seem natural to expect that $ \dinidl $ constitute a lower bound on the information loss $ \dinidd $ caused by optimal NI decoders. However, as we have recently shown, this relation need not hold \citep{eyherabide2013}. In this section, we begin our quest to disentangle whether this result indicates that the measures are flawed or that traditional expectations are unjustified, by comparing the information notions underlying $ \dinidl$ and $ \dinidd $.

\begin{figure}[htb!]
	\centering
	\includegraphics{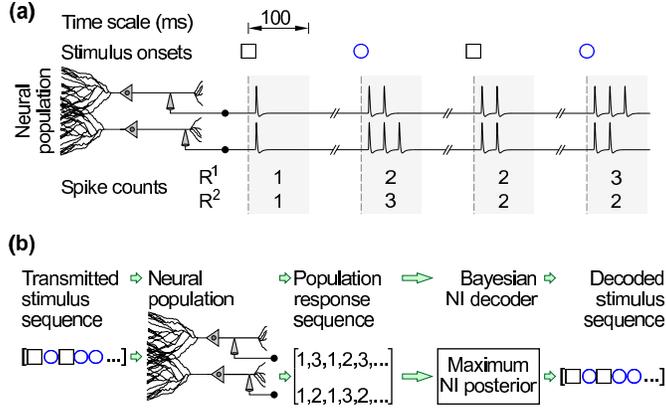}
	\caption[]{Communication notions previously used to derive $ \dinidl $. (a) Hypothetical intracellular recording of the concurrent spike trains produced by two neurons within $ 100\,ms  $ after presenting each of the visual stimuli $ \SBox $ and $ \SCirc $ in alternation, with pauses between each presentation. Only two types of population responses occur for each stimulus, and can be fully described using spike counts. (b) Representation of the memoryless communication system employed by \citet{latham2005} to compute $ \dinidl $.}\label{fig01}
\end{figure}

The measure $ \dinidl $ was derived by \citet{latham2005} using a notion of information with roots in communication theory, here called communication information (\sref{met::sec::neuenc}). Within the context of this study, this notion and the derivation of $ \dinidl $ can both be intuitively described using the population of two neurons depicted in \fref{fig01}(a). These two neurons can distinguish between two visual stimuli, namely $ \SBox $ and $ \SCirc $, by concurrently firing the same number of spikes ($ 1 $ or $ 2 $) after observing $ \SBox $, and different number of spikes ($ 2 $ or $ 3 $) after observing $ \SCirc $.

To compute $ \dinidl $, \citet{latham2005} employed sequences $ \MSEnc{=}[\MSEnc[1],\ldots,\MSEnc[N]] $ of $ N $ independent and identically distributed stimuli, and grouped into sets called codebooks (\fref{fig01}(b)). Each $ \MSEnc $ is transformed by the aforementioned neural population, one stimulus at a time and independently of the others, into a sequence $ \MR{=}[\MR[1],\ldots,\MR[N]] $ of $ N $ population responses. The $ \MR $ are then fed into an optimal NI decoder that attempts to reconstruct the original $ \MSEnc $ (\sref{met::sec::neudec}).

Using this interpretation, \citet{latham2005} computed an estimate of the average probability $ P(\MSEnc{\neq}\MSDec) $ that the decoded sequences $ \MSDec$ produced by optimal NI decoders differ from the transmitted sequences $ \MSEnc $, usually denoted $ \PSNIerr $. They showed this estimate to decay exponentially as $ N $ grows for codebooks of up to $ \exp[N\,\tilde{I}] $ sequences, from which they concluded, based on standard information-theoretical results, that $ \tilde{I} $ quantifies the communicated information. Subtracting $ \tilde{I} $ from the encoded information yielded the measure $ \dinidl $ that is currently believed the exact information loss caused when ignoring noise correlations in optimal decoding (\sref{met::sec::measures}).

Current beliefs notwithstanding, we have recently shown that $ \dinidl $ can exceed the information loss $ \dinidd $ caused by optimal decoders constructed ignoring noise correlations \citep[\sref{met::sec::measures};][]{eyherabide2013}. Analogously to previous conclusions about $ \dinid $, this finding may seem to directly indicate that $ \dinidl $ is flawed and overestimates the information loss caused by ignoring noise correlations in optimal decoding. However, as we note here, this conclusion overlooks the fact that $ \dinidd $ is fundamentally different from $ \dinidl $, and therefore not necessarily comparable.

Specifically, $ \dinidd $ differs from $ \dinidl $ in at least the following three fundamental aspects. First, it stems from treating decoding as a recoding process, rather than as the final stage in a communication system \citep{quiroga2009}. Second, it is sensitive to random errors but insensitive to systematic errors, and hence be large without decoders ever being correct \citep{schneidman2003,quiroga2009}. Third, its derivation involves single stimuli, as opposed to stimulus sequence and asymptotic limits. These three differences are not necessarily unknown in the neuroscience literature, but they are often overlooked.

Overlooking the differences between $ \dinidd $ and $ \dinidl $, and therefore deeming them comparable, may seem justified for at least two reasons. First, previous studies have shown that $ \dinid $ \citep{nirenberg2003,latham2013}, and consequently $ \dinidl $ \citep{eyherabide2016b}, reduces to the traditional information loss when applied to ignoring response aspects through transformations of the population response. Second, the data processing inequality ensures that, under those cases, $ \dinidl $ is a lower bound of the information loss caused by decoders that operates on the transformed responses. However, as we have recently shown, the second reason is invalid because, within the decoding perspective, decoders operate on the original responses, as opposed to the transformed ones. Furthermore, we have recently proved that, contrary to previously thought, the first reason need not be valid even for deterministic transformations \citep{eyherabide2016b}.

Therefore, we find questionable both to overlook the differences between the information notions underlying $ \dinidl $ and $ \dinidd $, and to draw conclusions conclusion about flaws in $ \dinidl $ based on previous observations that $ \dinidl{>}\dinidd $. Instead, those observations may indicate major departures from traditional relations between their underlying information notions. Nevertheless, it seems to us rather unfortunate that a putative exact measure of information loss cannot be regarded as a yardstick against which the performance of optimal NI decoders can be measured.

Resolving this puzzle requires that we hereinafter undertake at least two actions. First, distinguishing the notion of communication information underlying $ \dinidl $, from the notion of axiomatic information underlying $ \dinidd $, here so-called because it seemingly interprets information as an abstract measure of arbitrary correlations that fulfills certain intuitive and desirable axioms \citep{shannon1949,woodward1952,fano1961,gallager1968}. Second, assessing the properties of these two measures within the boundaries of their underlying information notions. In this way, our strategy becomes unfortunately more complex than those followed by previous studies, but guarantees to avoid their potential confounds, and to accurately compare $ \dinidd $ and $ \dinidl $.

\subsection{$ \dinidl $ overestimates the communication information loss}\label{res::sec::paradox}

We have recently shown not only that $ \dinidl $ can exceed $ \dinidd $, but also that $ \dinidl $ can be positive even when optimal NI decoders never make mistakes \citep{eyherabide2013}. However, the results there shown comprised a single hypothetical experiment analogous to the one depicted in \fref{fig01}(a), which may seem to possess peculiar characteristics. Most importantly, we did not connect those results with the possitibility of flaws in $ \dinidl $. In this section, we test this possibility, avoiding potential confounds by recasting the above experiment within the framework of \citet{latham2005}, and generalizing the results to any experiment in which optimal NI decoders never make mistakes, regardless of the stimulus-response distributions.

Imagine generic experiments in which optimal NI decoders can perfectly identify the $ \VS $ that elicited each $ \VR $, including and beyond that in \fref{fig01}(a). Mathematically, this means that $ P(\VS{\neq}\VSD){=}0 $, where $ \VSD $ denotes the decoded stimuli. As we note here, all these experiments can be recast within the framework of \citet{latham2005}, by interpreting neural populations as memoryless channels that read unit-length stimulus sequences $ \MSEnc[1] $. These sequences are turned into unit-length population-response sequences $ \MR[1] $, which are subsequently fed into optimal NI decoders.

Suppose now that the length of the sequences is increased to an arbitrary value $ N $. According to $ \dinidl $, optimal NI decoders may start making mistakes. Indeed, suppose that, in the experiment of \fref{fig01}(a), the population responses $ \VR{=}[R^1,R^2] $ associated with each stimulus are equally-likely and $ \SBox $ occurs more frequently than $ \SCirc $. Mathematically, this implies that $ P(\VR|\VS){=}0.5 $ regardless of $ \VR $ and $ \VS $, and that $ P(\SBox){>}0.5 $. In that case, $ \dinidl $ is positive and can reach ${\approx}25\,\% $ of the transmitted information \citep{eyherabide2013}, both of which can only occur if $ \PSNIerr{>}0 $.

To prove that this is not the case, recall that by hypothesis, optimal NI decoders can produce stimulus estimates $ \MSDec[n] $ for each $ \nth $ population response $ \MR[n] $ within the received  $ \MR $ without errors, namely $ P(\MSEnc[n]{\neq}\MSDec[n]){=}0 $. Boole's inequality \citep{casella2002} ensures that concatenating these parallel estimates to produce the decoded stimulus sequence $ \MSDec $ yields no sequence-errors, that is

\begin{equation}
P(\MSEnc{\neq}\MSDec)\leq N\,\max_{n}{P(\MSEnc[n]{\neq}\MSDec[n])} = 0\, .
\end{equation}

\noindent Following \citet{latham2005}, this result implies that, in the above experiments, ignoring noise correlations in optimal decoding never causes communication information losses.

Therefore, we have proved for the first time that $ \dinidl $ is flawed and overestimates the communication information loss when ignoring noise correlations in optimal decoding. For this reason, we denote the latter as $ \Delta CI $. Momentarily, we will define $ \Delta CI $ as follows

\begin{equation}
\Delta CI = \begin{cases}0 & \mbox{in the absence of decoding errors}\\ \dinidl & \mbox{otherwise} \end{cases}\, ,
\end{equation} 

\noindent but refine it later after revealing additional sources of overestimation.

\subsection{Misleading intuitions and limitations}\label{res::sec::misleading}

Although we proved above that $ \dinidl $ overestimates $ \Delta CI $, our proof still partially rests on the experiment in \fref{fig01}(a). This experiment possesses three properties that may seem peculiar and in direct contradiction with current intuitions about the role of noise correlations in optimal decoding. In this section, we show these properties inessential for $ \dinidl $ to overestimate $ \Delta CI $.

Our demonstrations focus on cases in which optimal NI decoders make no mistakes. This property implies that each $ \VR $ occurs for only one $ \VS $. However, we exclude the cases in which each $ \VR $ would occur for only one $ \VS $ should neurons be NI, because the resulting $ \dinidl $ is trivially zero and therefore tight. In the remaining cases, at least one $ \VR $ that occurs when neurons are correlated always occur for more than one $ \VS $ should neurons be NI.

\begin{figure}[htb!]
	\centering
	\includegraphics{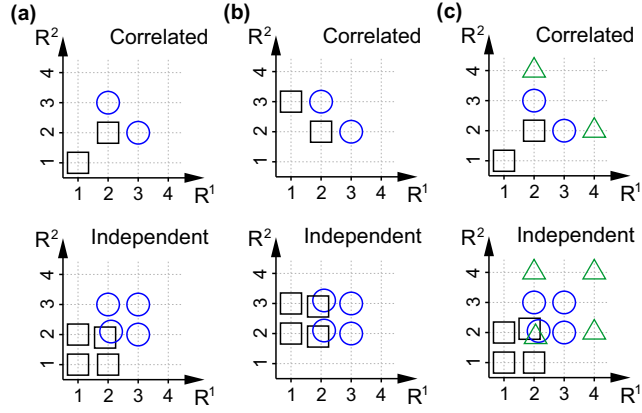}
	\caption[]{$ \dinidl $ may overestimate the communication information loss. (a) Cartesian representation of the responses recorded in the experiment of \fref{fig01}(a) (top) and the surrogate NI responses that would have occur should neurons be NI (bottom). (b) Analogous description to panel (a), but for the responses recorded in another hypothetical experiment. (c) Analogous description to panel (b). In all cases, the symbols represent the stimuli; and their locations, the concurrent responses they elicit. To improve visibility, symbols that occur in the same location are shown to only partially overlap.}\label{fig02}
\end{figure}

To illustrate this condition, we constructed three experiments analogous to \fref{fig01}(a) and represented them using Cartesian coordinates (\fref{fig02}). In all panels, the $ \VR $s recorded during the experiment are represented at the top, and those that would have been recorded should neurons be NI are represented at the bottom. The experiment in \fref{fig02}(a) actually coincides with that in \fref{fig01}(a), whereas those in \fref{fig02}(b) and \fref{fig02}(c) are variations of it explained below. In all cases, the Cartesian representations clearly show that each $ \VR $ occurs for only one $ \VS $ when neurons are correlated (top) and that $ \VR{=}[2,2] $ would have occurred for all $ \VS $s should neurons be NI.

The first seemingly peculiar property of \fref{fig01}(a) is that noise correlations vary across $ \VS $. Indeed, \fref{fig02}(a) shows that the concurrent responses of both neurons are positively correlated when elicited by $ \SBox $ (that is, the Pearson correlation coefficient $ \rho(R^1,R^2|\SBox){>}0 $), and negatively correlated when elicited by $ \SCirc $ (that is, $ \rho(R^1,R^2|\SCirc){<}0 $). This observation has previously been thought to indicate that noise correlation must be important from the decoding perspective \citep{nirenberg2001addmeister,latham2005}. However, in \sref{res::sec::paradox}, we proved that this intuition need not be correct.

Most importantly, this first property is insufficient and unnecessary for proving that $ \dinidl $ overestimates $ \Delta CI $. To prove it insufficient, we take the experiment in \fref{fig02}(a) and set $ P(\VR|\VS){\neq}0.5 $ regardless of $ \VS $ and $ \VR $, thereby obtaining through \eref{met::def::dinidl} that $ \dinidl{=}0 $ \citep{eyherabide2013}. To prove it unnecessary, we build another experiment analogous to \fref{fig02}(a), but with $ \VR{=}[1,1] $ replaced by $ [1,3] $ (\fref{fig02}(b)). Contrary to \fref{fig02}(a), here noise correlations remain constant across stimuli both in sign and in strength (i.e., $ \rho(R^1,R^2|\SBox){=}\rho(R^1,R^2|\SCirc){=}-1 $). Notwithstanding, setting $ P(\SBox){=}0.52$, $P(\VR{=}[2,2]|\SBox){=}0.48 $, and $ P(\VR{=}[2,3]|\SCirc){=}0.52 $, yields $ \dinidl{\approx}50\,\% $ even though optimal NI decoders can be readily proved to never make mistakes.

The second seemingly peculiar property is that the value of $ \dinidl$ equals the value of its predecessor, namely $\dinid $ \citep{eyherabide2013}. Interestingly, this observation has previously been thought to indicate that $ \dinid $ is tight \citep{latham2013}. On the contrary, here we show that such observation actually indicates that $ \dinidl $ is loose. Specifically, we can prove that $ \dinidl{=}\dinid $ is sufficient but inessential for proving that $ \dinidl $ overestimates the communication information loss. To avoid clutter, we hereinafter omit the response labels from the arguments of the probabilities.

To prove it sufficient, notice that, for the cases under consideration (see second paragraph), at least one $ \VR $ and $ \VS $ exists for which $ \PNIs(\VS|\VR){<}1 $ and $ P(\VS|\VR){=}1 $. Therefore, \eref{met::def::dinid} yields that $ \dinid{>}0 $, and consequently, the second property implies that $ \dinidl{>}0 $. However, the second property need not imply that optimal NI decoders make mistakes. Indeed, the property holds when setting $ P(\VS,\VR) $ in \fref{fig02}(a) so that $ \PNIs([2,2]|\SBox){=}\PNIs([2,2]|\SCirc) $, even though optimal NI decoders never make mistakes when $ P(\SBox){>}0.5 $.

To prove it inessential, we set in \fref{fig02}(b) $ P(\SBox){=}0.48 $, $ P([2,2]|\SBox){=}0.58 $, and $ P([2,3]|\SCirc){=}0.48 $. Using these probabilities, \eref{met::def::dinid} and \eref{met::def::dinidl} readily yield different values for $ \dinid$ (${\approx}48\,\% $) and $ \dinidl$ (${\approx}36\,\% $), respectively. However, optimal NI decoders can be readily proved to never make mistakes.

The third seemingly peculiar property is that each $ \VR $ that would occur for more than one $ \VS $ should neurons be NI, would do so with equal frequency for those $ \VS $. Mathematically, $ \PNIs(\VR|\VS_A){>}0 $ and $ \PNIs(\VR|\VS_B){>}0 $ implies that $ \PNIs(\VR|\VS_A){=}\PNIs(\VR|\VS_B) $. However, this property is sufficient but typically inessential for proving that $ \dinidl $ overestimates $ \Delta CI $.

We prove it sufficient by noting that this property turns $ \dinidlt{\VS;\VR|} $ independent of $ \theta $, and thus $ \dinidl{=}\dinid $. To prove it typically inessential, we first note that the third property is necessary for experiments comprising only two $ \VS $ and only one $ \VR $ for which $ \PNIs(\VR|\VS_A){=}\PNIs(\VR|\VS_B) $, because otherwise $ \dinidlt{\VS;\VR|}{\rightarrow }0 $ either when $ \theta{\rightarrow}{-}\infty $ or $ {+}\infty $, thereby yielding $ \dinidl{=}0 $.

However, this need not be the case when experiments comprise either more than one $ \VR $ for which $ \PNIs(\VR|\VS_A){=}\PNIs(\VR|\VS_B) $ even if for only two stimuli, as we have already shown during our proofs of the first and the second properties, or a single $ \VR $ with $ \PNIs(\VR|\VS_A){=}\PNIs(\VR|\VS_B) $ for more than two stimuli. To prove the latter, we take the experiment of \fref{fig02}(a) and add two population responses elicited by a third stimulus $ \STri $ (\fref{fig02}(c)). Setting $ P(\VR|\VS)$ so that $\PNIs([2,2]|\STri){<}\PNIs([2,2]|\SBox){<}\PNIs([2,2]|\SCirc) $ (or in reversed order) turns $ \dinidl{>}0 $ regardless of the stimulus probabilities, even though it can be readily proved that stimulus probabilities always exist for which optimal NI decoders never make mistakes.

To summarize, we have shown that the seemingly peculiar properties of the experiment in \fref{fig01}(a) are inessential for proving that $ \dinidl $ overestimates $ \Delta CI $. Contrary to previous beliefs, we have also found that the sign and strength of noise correlations can be misleading about their role in optimal decoding. Most importantly, we have proved that the proximity of $ \dinid $ to $ \dinidl $ need not indicate that $ \dinid $ is close to $ \Delta CI $, but that $ \dinidl $ is loose.

\subsection{Overestimation in the presence of decoding errors}\label{res::sec::overest}

\begin{figure}[htb!]
	\centering
	\includegraphics{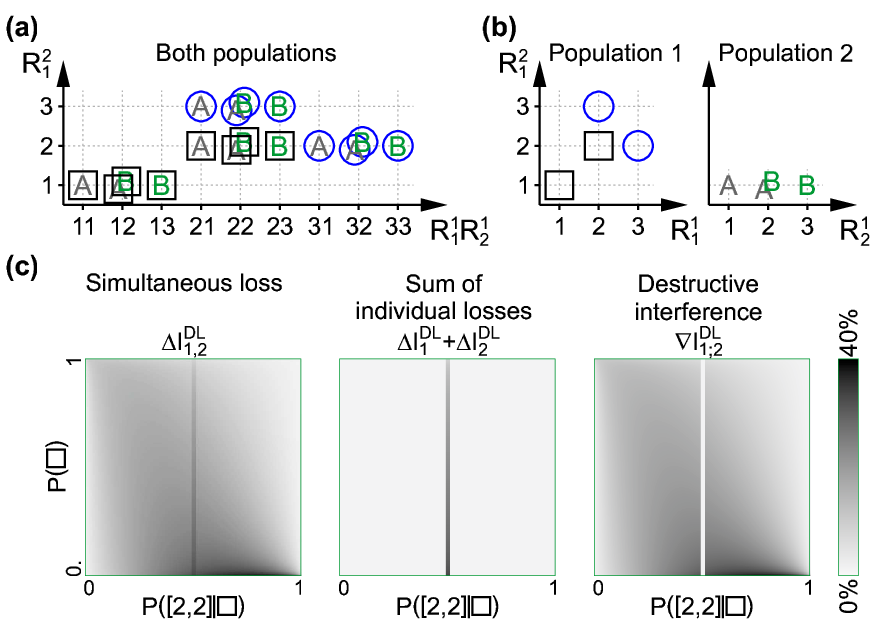}
	\caption[]{Correlations become paradoxically important when integrating independent information. (a) Cartesian representation of the concurrent responses $ \VR{=}[R_1^1,R_1^2,R_2^1] $ of three neurons elicited by four different compound stimuli ($ \SCA $, $ \SCB $, $ \SBA $ and $ \SBB $). (b) Cartesian representations of the concurrent responses of $R_1^1$ and $R_1^2 $ (left), which are sensitive only to variations in the frames; and the responses $ R_2^1 $ (right), which are only sensitive to variations in the letters. (c) Density or heat maps (with color-scale given on the right) depicting, as a function of $ P(\SBox) $ and $ P([2,2]|\SBox) $, the following three quantities: the value of $ \dinidl $ when decoding frames and letters together (left); the sum of the values of $ \dinidl $ when decoding them singly (middle); and the resulting destructive interference (right). The line $ P([2,2]|\SBox){=}0.5 $ has been widened to ease the visualization of discontinuities.}\label{fig03}
\end{figure}

Our analysis so far may seem limited to cases in which optimal NI decoder make no mistakes, which is arguably rare in the nervous system. In this section, we show that the measure $ \dinidl $ may overestimate $ \Delta CI $ even when optimal NI decoders do make mistakes. To prove this, consider the experiment shown in \fref{fig03}(a) depicting the concurrent responses $ \VR{=}[R_1^1,R_1^2,R_2^1] $ of three neurons elicited by four different $ \VS $, namely $ \SCA $, $ \SCB $, $ \SBA $ and $ \SBB $.

In this experiment, $ \VS{=}[S_1,S_2] $ consist of two independently chosen features: a frame ($ S_1{\in}\{\SBox,\SCirc\} $) and a letter ($ S_2{\in}\{\SA,\SB\} $). The neurons can be separated into two populations that fire independently of one another and selectively to different stimulus features (\fref{fig03}(b)). The first population is only sensitive to the frames and analogous to that in \fref{fig01}(a). The second population is only sensitive to the letters, and consists of a single neuron. As a result, each population constitutes an independent information stream, namely $ P(\VS,\VR){=}P(S_1,\VR_1)\,P(S_2,\VR_2) $ (\sref{met::sec::indinfo}).

Suppose that we set for the second population $ P(2|\SA){=}P(2|\SB)$. In that case, \eref{met::def::dinidl} readily yields that the value of $ \dinidl $ computed using all neurons equals the one computed using only the first two. However, these two neurons are analogous to those in \fref{fig01}(a), for which we have already shown that $ \dinidl $ may overestimate $ \Delta CI $. Therefore, it seems evident that $ \dinidl $ also overestimates $ \Delta CI $ for the experiment in \fref{fig03}(a) as well.

Using this result, we can refine the computation of $ \Delta CI $ as follows

\begin{equation}
\Delta CI \begin{cases}=0 & \mbox{if $ P(\VS{\neq}\VSD){=}0 $}\\ 
\leq\dinidl & \mbox{otherwise} \end{cases}\, ,
\end{equation} 

\noindent thereby reflecting our findings that, contrary to previous studies, $\dinidl $ cannot be ensured tight. Although this conclusion is correct, the rationale need not be general and should therefore be observed with caution. The reason for our concerns lies in paradoxical properties of $ \dinidl $ that, as we reveal in the next section, emerge when setting $ P(\VS,\VR) $ to other values than those used in our demonstration.

\subsection{The whole is less than the sum of its independent parts}\label{res::sec::together}

We have just shown for the first time that $ \dinidl $ may overestimate $ \Delta CI $, but the overestimation may seem to occur only when at least some stimulus features (as opposed to stimulus identities) can be perfectly identified. These cases are arguably rare in the nervous system where stimulus features can almost never be perfectly identified due to noise, in which case $ \dinidl $ may still be exact. Surprisingly, here we show that, in those cases, the value of $ \dinidl $ computed over all neural populations typically exceeds the sum of the values of $ \dinidl $ computed over each neural population, even when neural populations fire independently and selectively to different and independent stimulus features, thereby transmitting independent information (section~\ref{met::sec::indinfo}).

Specifically, suppose that in \fref{fig03}(a) we set $ P(\SB)$ equal to $P(\SBox)$. In addition, we set $ P(1|\SA)$,  $P(2|\SB)$, and $P([2,3]|\SCirc)$, equal to  $P([2,2]|\SBox) $, where without risk of confusion we have omitted the response labels $ R_j^k $ within the arguments of the probabilities to avoid clutter. Under these conditions, the computation of $ \dinidl $ using all neurons in both populations yields a value here denoted $ \dinidl_{1,2} $ given by the following

\begin{equation}\label{res::eq::dinidl12fig03}
\dinidl_{1,2} = \begin{cases}-0.5\,q\,\ln q \hspace{-3ex}& \mbox{if $ \alpha {=} 0.5 $} \\ \phi \,\Hopb(\qa/\phi) - \qa\,\ln 4 \hspace{-3ex}& \mbox{otherwise},\end{cases}
\end{equation}

\noindent where for compactness, we have defined $ q{=}P(\SBox) $, $ \alpha{=}P([2,2]|\SBox) $, $ \Hopb(x){=}-x \ln x{-}{\bar{x}}\ln{\bar{x}}$ (also called binary entropy function), $ \phi{=}2\, \qa+\qcac $, and the bar over a symbol denotes complement to unity (e.g., $ \bar{z}{=}1{-}z $).

Notice that Pearson correlation coefficients ($ \rho $) cannot fully characterize the response distributions. Indeed, the only nonzero $ \rho $s arise for the responses of both neurons in population 1. For $ \SBA $ and $ \SBB $, $ \rho{=}1 $, whereas for $ \SCA $ and $ \SCB $, $ \rho{=}{-}1 $, and their values remain constant regardless of $P([2,2]|\SBox) $. In addition, $ \dinidl $ depends on $P([2,2]|\SBox) $, as we show in \eref{res::eq::dinidl12fig03} and \fref{fig03}(c), thereby rendering $ \rho $ also unsuitable for assessing the properties of $ \dinidl $.

The value of $ \dinidl_{1,2} $ in \fref{fig03}(a) is almost always positive, and can reach ${\approx}100\,\% $ of the transmitted information when $ P(\SBox){\rightarrow}0$ and $P([2,2]|\SBox){\rightarrow}1$ (\fref{fig03}(c) left, and \ref{se::fig1}). According to previous studies, this result would indicate that noise correlations are important in optimal decoding or even crucial \citep{latham2005,ince2010,oizumi2010,latham2013,oizumi2016}. However, we find the above result paradoxical because noise correlations only exist within population 1, and, unless $ P([2,2]|\SBox) {=}0.5$, they are irrelevant for decoding that population \citep{eyherabide2013}.

Indeed, the computation of $ \dinidl $ using only the neurons in the first population always yields $ \dinidl_1{=}0 $, unless $ P([2,2]|\SBox){=}0.5 $, whereas its computation using only the neurons in the second population always yields $ \dinidl_2{=}0 $, regardless of $ P([2,2]|\SBox) $ (\fref{fig03}(c) middle). Therefore, the positive values of $ \dinidl_{1,2} $ seem to arise from a paradoxical growth in correlation importance caused by decoding the populations together, as opposed to decoding them in parallel, even though the populations transmit independent information. This paradoxical growth is here called destructive interference, and quantified as the following difference

\begin{equation} \label{res::eq::interference}
\dintdl_{1;2} = \dinidl_{1,2} - \dinidl_{1} - \dinidl_{2}\, . 
\end{equation}

\noindent In our experiment, $ \dintdl_{1;2}$ is always positive and equal to $ \dinidl_{1,2} $ (though zero when $ P([2,2]|\SBox){=}0.5 $), and therefore can reach $ {\approx}100\,\% $ of the transmitted information (\fref{fig03}(c) right).

Notice that the above results have not been corrected for the overestimation problems described in \sref{res::sec::paradox} and \sref{res::sec::overest}. As a result, the observed destructive interference cannot be directly attributed to an overestimation of the types there studied. Instead, it constitutes a new phenomena that can occur in parallel with other sources of overestimation.

Indeed, as shown in \sref{res::sec::paradox}, the actual communication information loss $ \Delta CI_1 $ computed using only the neurons in the first population is equal to zero when optimal NI decoders make no mistakes, and this can occur even if $ P([2,2]|\SBox){=}0.5 $. To that end, one of the following two conditions must hold: $ P(\SBox){>}0.5 $, or $ P(\SBox){=}0.5 $ but optimal NI decoders are constructed using tie-breaking rules that choose $ \SBox $ whenever $ \PNIs(\SBox|\VR_1){=}\PNIs(\SCirc|\VR_1) $ \citep{eyherabide2013}. Unfortunately, we can not refine the estimation of the actual communication information loss $ \Delta CI_{1,2} $ computed using all neurons in all populations based on the results in \sref{res::sec::overest} because, unlike the example there studied, here $ \dinidl_{1,2}{>}\dinidl_1 $. This result would imply that the actual destructive interference is zero only when $ P([2,2]|\SBox){=}0.5 $ and $ P(\SBox){<}0.5 $, that is, on the lower half of the line defined by $ P([2,2]|\SBox){=}0.5 $ in \fref{fig03}(c).

In conclusion, here we have shown for the first time that $ \dinidl $ is superadditive when information is independent. This result is in stark contrast with the additivity of its predecessor, $ \dinid $, which we later prove in \sref{res::sec::ubiquity}. Most importantly, our observation reveals a major departure from traditional views on the desirable properties of information measures \citep{fano1961,cover2006}. From a neuroscientific perspective, our result implies that, should $ \dinidl $ be exact as currently thought, noise correlations would paradoxically become more important when analyzing neural populations as a whole than when analyzing each of its constituent parts, even when the parts transmit independent information.

\subsection{Destructive interference in brain models}\label{res::sec::brain}

Information in the brain may be transmitted by one or multiple heterogeneous neural substrates (e.g. single neurons, neural populations, cortical areas or functional networks) using different temporal scales, frequency bands, amplitude intervals, or other types of multiplexed codes \citep{shannon1949,oppenheim1997,cover2006,eyherabide2010FCN,panzeri2010,huk2012,gross2013,harvey2013,akam2014}. Moreover, neural activity may be characterized using continuous (quantitative) variables, as opposed to discrete (qualitative, nominal or categorical) variables. Continuous variables naturally arise when estimating firing rates, peak amplitudes, phases, or mean power within specific frequency-bands and time-intervals using tuning curves, local-field potentials (LFP), event-related potentials (ERP), or sensor signals from brain-imaging devices including electroencephalography (EEG) and magnetoencephalography (MEG). Unless quantized (discretized), these estimations may yield locally-smooth probability densities for which the above paradox need not arise.

\begin{figure}[htb!]
	\centering
	\includegraphics{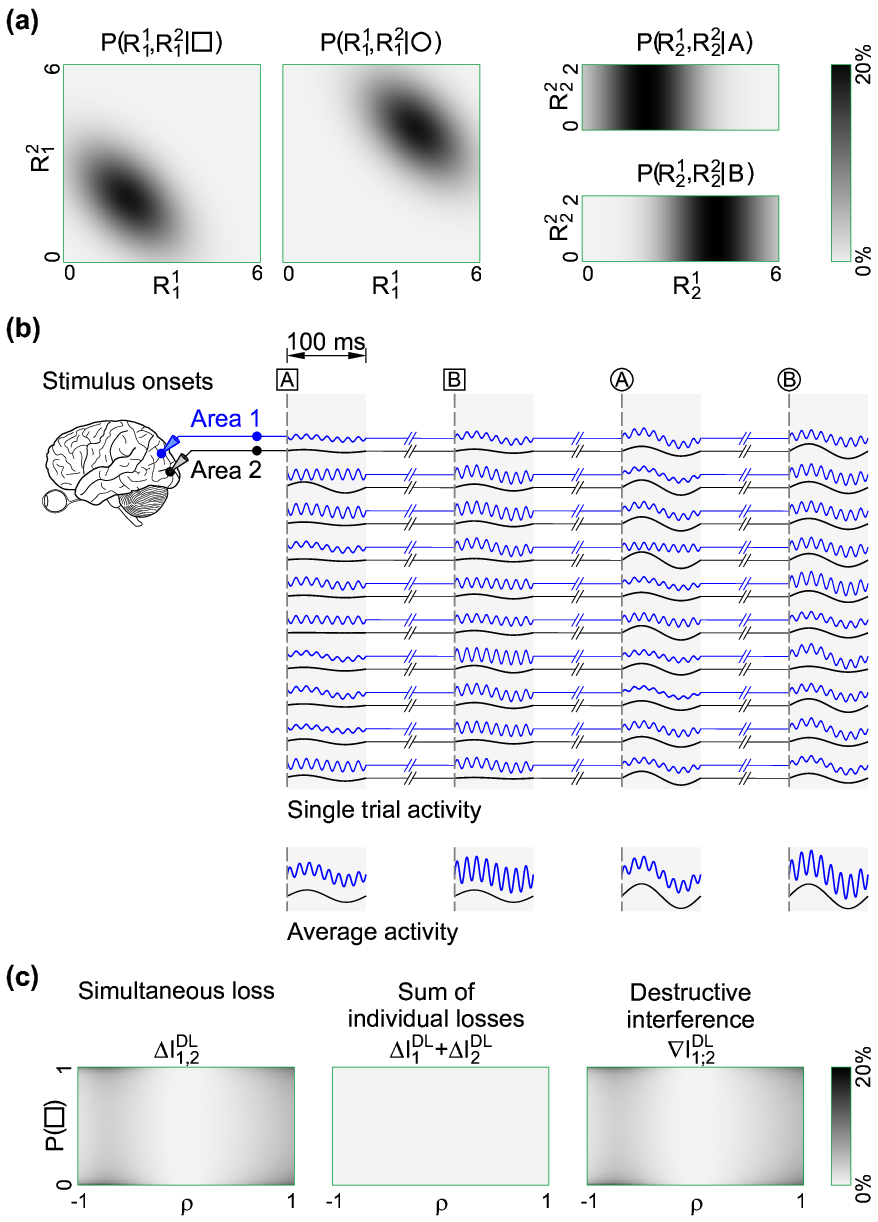}
	\caption[]{Destructive interference in brain models. (a) Cartesian representation of the probability densities defined in \eref{res::eq::brainprobabilities1} and \eref{res::eq::brainprobabilities2}. The dummy variable $ R_2^2 $ uniformly distributed within the interval $ [0,2] $ was introduced for visualization purposes. (b) Hypothetical recording of the activity in two cortical areas, elicited within $ 100\,ms $ after the onset of each stimulus $ \SBA $, $ \SBB $, $ \SCA $ and $ \SCB $, and defined through \eref{res::eq::brainactivity1} and \eref{res::eq::brainactivity2}. (c) Analogous description to \fref{fig03}(c), but with $ P([2,2]|\SBox) $ replaced by $ \rho $.}\label{fig04}
\end{figure}

To test this, we build another hypothetical experiment by replacing the discrete responses in \fref{fig03}(a) with continuous ones generated using unit-variance Gaussian distributions (\fref{fig04}(a)). Mathematically, their probability distributions can be rewritten as follows

\begin{align}
P(\VR_1|S_1) &\propto \exp\lrcor{\frac{\big(R_1^1-R_1^2\big)^2}{2\,({\rho}^2-1)}-\frac{(R_1^1-\mu_1)\,(R_1^2-\mu_1)}{1+\rho}}\label{res::eq::brainprobabilities1}\\
P(\VR_2|S_2) &\propto \exp\lrcor{-0.5\,\big(R_2^1-\mu_2\big)^2}\, .\label{res::eq::brainprobabilities2}
\end{align}

\noindent where $ \rho $ is the correlation coefficient between $ R_1^1 $ and $ R_1^2 $; $ \mu_1 {=} 2$ if $ S_1 {=} \SBox $ and $ 4$ if $ S_1 {=} \SCirc$; and $ \mu_2 {=} 2$ if $ S_2 {=} \SA $ and $ 4$ if $ S_2 {=} \SB$.

To study multiplexed codes, we reinterpret $ R_1^1 $ and $ R_1^2 $ as the amplitudes of $ 10\,Hz $-oscillations at two different cortical areas, denoted $ 1 $ and $ 2 $. Analogously, $ R_2^1 $ as the amplitude of $ 80\,Hz $-oscillations at the cortical area $ 1 $. Mathematically, 

\begin{align}
Z_1 &= R_1^1\,\sin(20\,\pi\,t) + R_2^1\,\sin(160\,\pi\,t)\label{res::eq::brainactivity1}\\
Z_2 &= R_1^2\,\sin(20\,\pi\,t)\, ,\label{res::eq::brainactivity2}
\end{align}

\noindent where $ Z_1 $ and $ Z_2 $ represent the possibly-filtered signals recorded from the cortical areas $ 1 $ and $ 2 $, and $ t $ is the time (\fref{fig04}(b)).

Analogous to the experiment in the previous section, noise correlations only occur between $ R_1^1 $ and $ R_1^2 $. However, here they affect the amplitudes of the $ 10\,Hz $-oscillations at the two different cortical areas, as opposed to the spike counts of two neurons within the same population. Nevertheless, they need not be important from the decoding perspective regardless of their sign and strength, as previous studies have already shown \citep{averbeck2006,averbeck2006b,eyherabide2013}.

Unfortunately, the results and conclusions of the above studies cannot be directly applied to our experiment because they were obtained using different measures and information notions. Nevertheless, we can rigorously prove that value of $ \dinidl $ computed taking only the $10\,Hz $-oscillations into account, here denoted $ \dinidl_{1} $, is zero regardless of $ \rho $. Indeed, after some algebra, $ P(\VS|[R_1^1,R_1^2]) $ can be proved equal to $ \PNIs(\VS|[R_1^1,R_1^2],\theta) $ for $ \theta{=}1/(1+\rho) $ whenever $ P([R_1^1,R_1^2]){>}0 $.

According to previous studies, this result would indicate that noise correlations are irrelevant for decoding them in isolation regardless of the correlation strength. Even so, the value of $ \dinidl_{1,2} $ computed using all oscillations amplitudes in all cortical areas is always positive (except when $ \rho{=}0 $). Based on our results in the previous section, we can conclude that $ \dinidl_{1.2} $ is completely attributable to destructive interference, namely $ \dinidl_{1,2}{=}\dintdl_{1;2} $ (\fref{fig04}(c)).

In conclusion, we have shown that destructive interference is ubiquitous regardless of the type of variables employed to characterize the recorded neural activity, and the neural substrates that the variables are interpreted to represent. In this way, we have conclusively answered two of the most recurrent questions in computational neuroscience: whether the phenomena observed when studying single neurons would also occur when studying cortical areas, and whether the results obtained with discrete variables would also emerge for continuous variables.

In addition, recall that we interpreted \fref{fig04} as an example of multiplexed codes that transmit independent information through frequency division, whereas we can interpret \fref{fig03} as an example of multiplexed codes that transmit independent information through space division \citep{oppenheim1997,panzeri2010}. These interpretations allow us to predict that destructive interference is a characteristic feature of multiplexed codes regardless of their implementation. However, our prediction need not imply that demultiplexing improves the performance of optimal NI decoders when information is independent, as we show in \sref{res::sec::moreefficient}.

\subsection{Ubiquity of destructive interference}\label{res::sec::ubiquity}

In the previous section, we quantified $ \dinidl $ using only two populations transmitting independent information and found that it is superadditive. This result constitutes a major departure both from traditional views on the desirable properties of information measures, and from traditional expectations when operating on independent-information streams \citep{fano1961,schneidman2003,cover2006,oizumi2016}. However, the aforementioned experiments may still seem overly simple, thereby questioning the generality of our results. In this section, we trace back this phenomenon to the mathematical definition of $ \dinidl $, as opposed to particular properties of the hypothetical neural data analyzed above.

Recall that $ \dinidl $ is defined through \eref{met::def::dinidl} as the minimization over the parameter $ \theta $ of the function $ \dinidlt{} $, which is given by the following

\begin{equation}\label{res::eq::dinidlt}
\dinidlt{\VS;\VR|} = \sum_{S,\VR}{P(S,\VR)\,\ln \frac{\PSdR}{\PNIs\sdrtharg}}\, .
\end{equation}

\noindent When two neural populations transmit independent information, like those in \fref{fig03} and \fref{fig04}, it can be readily shown that, in addition to \eref{met::eq::indjoint}, the following equations hold 

\begin{align} 
\PSdR & = P(S_1|\VR_1)\,P(S_2|\VR_2)\label{met::eq::posteriors1}\\
\PNIs\sdrtharg &= \PNIs(S_1|\VR_1,\theta)\,\PNIs(S_2|\VR_2,\theta)\label{met::eq::posteriors3}\, ,
\end{align}

\noindent where $ S_j $ denotes the stimulus features encoded in the activity $ \VR_j $ of the $ \jth $ population. Therefore, we can rewrite \eref{res::eq::dinidlt12} for these cases as follows 

\begin{equation}\label{res::eq::dinidlt12}
\dinidlt{\VS;\VR|} = \dinidlt{\VS_1;\VR_1|} + \dinidlt{\VS_2;\VR_2|}\, ,
\end{equation}

\noindent thereby proving that $ \dinidlt{} $ is additive when information is independent. For compactness, hereinafter we employ the abbreviated notation introduced in \sref{met::sec::indinfo},  and rewrite the above equation simply as $ \dinidltsub{}{1,2}{=}\dinidltsub{}{1}+\dinidltsub{}{2} $.

Notice that the additivity of $ \dinidlt{} $ directly implies the additivity of $ \dinid $. Indeed $ \dinid $ can be computed as $ \dinidlt{} $ with $ \theta{=}1 $ \citep{latham2005}. Therefore, using the aforementioned notation, $ \dinid_\sj{=}\dinid_\spj $. This result should not be confused with the additivity found in \citet{nirenberg2003}, which involved neither decoders nor independent information, and is limited to non-overlapping response distributions.

However, the additivity of $ \dinidlt{} $ need not imply the additivity of $ \dinidl $. Indeed, $ \dinidlt{} $ is convex \citep[also called U-concave][]{gallager1968,latham2005}, and the sum of the minima to two convex functions can never exceed the minimum of their sum. Therefore, the minimum of each tern in \eref{res::eq::dinidlt12}, which correspond from left to right to $ \dinidl_{1,2} $, $ \dinidl_1 $ and $ \dinidl_2 $, are related according to the following

\begin{equation}\label{se::eq::dinidlgreater12}
\dinidl_{1,2} \geq \dinidl_1 + \dinidl_2\, .
\end{equation}

\begin{figure}[htb!]
	\centering
	\includegraphics{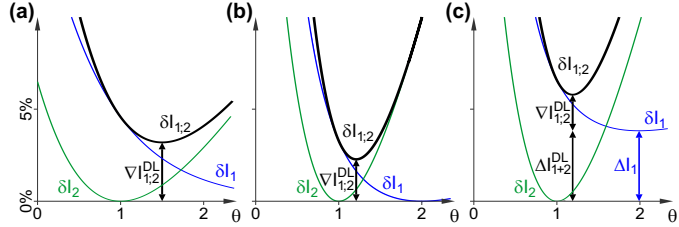}
	\caption[]{Destructive interference stems from differences in the location of minima.}\label{fig05}
\end{figure}

\noindent Strict inequality holds whenever the minima occur at different locations, irrespective of their separation or the values of the minima (\fref{fig05}).

To illustrate this, we build the three examples shown in \fref{fig05}. The first example is based on \fref{fig03}(a) with $ P(\SBox)$ and $P([2,2]|\SBox)$ equal to $0.8 $ (\fref{fig05}(a)), and shows the characteristic U-shape of $ \dinidlt{} $ regardless of whether it is computed using the neurons in population 1, in population 2, or in both. The minimum of $ \dinidltsub{}{1} $, namely $ \dinidl_1 $, occurs at $ \theta{\rightarrow}\infty $, whereas the minimum of $ \dinidltsub{}{2} $, namely $ \dinidl_2 $, occurs at $ \theta{=}1 $. As expected, their sum is less than the minimum of $ \dinidltsub{}{1,2} $, namely $ \dinidl_{1,2} $, thereby leading to destructive interference.

The second example is based on \fref{fig04}(a) with $ P(\SBox){=}0.8 $ and $ \rho{=}{-0.5} $ (\fref{fig05}(b)). This example is qualitatively analogous to the first one except that $ \dinidl_1 $ occurs not at $ \infty $, but at $ \theta{=}2$. The third example (\fref{fig05}(c)) is analogous to the second one, but with different correlation coefficients for $ \SBox $ ($ -0.8 $) and $ \SCirc $ ($ -0.2 $). As a result, $ \dinidl_1 $ is not zero, but positive and equal to $ {\approx}3.8\,\% $. In both cases, the observed differences are inessential and cannot prevent the occurrence of destructive interference.

Nevertheless, notice that a value of $ \theta $ that simultaneously minimizes each and every term in \eref{res::eq::dinidlt12} may theoretically occur. Indeed, recall the experiment in \fref{fig03} with $ P(\VS,\VR) $ set as in \sref{res::sec::overest}. In that case, we found that $ \dinidl_2{=}0 $ and $ \dinidl_{1,2}{=}\dinidl_1 $, from which we can immediately conclude that $ \dintdl_{1;2}{=}0 $. This conclusion agrees with the fact that both $ \dinidltsub{}{1} $ and $ \dinidltsub{}{2} $ are both minimized for the same value of $ \theta $. The fact that the value of $ \theta $ turns out to be unity is inessential.

Our results not only prove that the superadditivity of $ \dinidl $ stems directly from its mathematical definition, but also that the property is ubiquitous and independent of the type of data on which $ \dinidl $ is applied. Most importantly, they show that previous experimental findings in which $ \dinidl $ either grows with the number of neurons or with the decoding-window length or lies close to $ \dinid $, need not be completely attributable to temporal correlations across time bins, pseudo-correlations caused by inappropriately assuming stationarity, or higher-order correlations, as previous studies have conjectured \citep{oizumi2009,oizumi2010,latham2013}. Instead, they can at least partially arise, even when information is independent, due to destructive interference.

Answering these questions requires that we test our results on arbitrary number $ J $ of neural populations. To that end, we use mathematical induction and rewrite $ \dinidlj $ as the minimization of the convex function $\dinidltsub{}{\sj}{=}\dinidltsub{}{1,\ldots,J{-}1}{+}\dinidltsub{}{J} $, which minima are related through $ \dinidlj{\geq}\dinidl_{1,\ldots,J{-}1}{+}\dinidl_J $. This result implies that 

\begin{equation}\label{se::eq::dinidlgreater}
\dinidlj \geq \dinidlpj\, ,
\end{equation}

\noindent with equality if and only if some $ \hat{\theta} $ exists that simultaneously minimizes $ \dinidltsub{}{j} $ for all $ 1{\leq}j{\leq}J $ (that is, $ \dinitdl_j(\hat{\theta}){=}\min_{\theta}{\dinidltsub{}{j}} $ for all $ 1{\leq}j{\leq}J $).

Based on the above result, we can define destructive interference for an arbitrary number $ J $ of independent information streams as the following difference 

\begin{equation}
\dintdlj = \dinidlj - \dinidlpj\, . 
\end{equation}

\noindent It immediately follows from our demonstration of \eref{se::eq::dinidlgreater} that $ \dintdlj $ never decreases with $ J $.

In addition, the condition for equality in \eref{se::eq::dinidlgreater}
immediately implies that the observation of $ \dintdlj{>}0 $ requires of two independent information streams $ j_1 $ and $ j_2 $, for which $ \dinidltsub{}{j_1} $ and $ \dinidltsub{}{j_2} $ are strictly convex, and achieve their minima at different values of $ \theta $. These conditions need not always hold, for $ \dinidlt{} $ can also be constant \citep{eyherabide2013}, as in the cases that fulfill the third property mentioned in \sref{res::sec::misleading}. However, the fact that in these cases $ \dinidl{=}\dinid $ need not imply that this condition should not hold for both streams. Indeed, destructive interference can arise even if only one stream exists for which $ \dinidl{\neq}\dinid $, as shown in \fref{fig05}.

In conclusion, we have shown that destructive interference is a direct consequence of the convex minimization that defines the estimate $ \dinidl $ of the actual communication information loss $ \Delta CI $. In addition, we have derived the necessary and sufficient conditions for destructive interference to arise. These conditions need not always hold, as in \fref{fig03} with the probabilities set as in \sref{res::sec::overest}, but our results show that the conditions are quite unrestrictive. Most importantly, we have extended the validity of our results to arbitrarily-complex independent information streams, regardless of their number and type.

\subsection{Relative monotonic growth}\label{res::sec::morestreams}

We have just shown that $ \dintdlj $ never decreases with the number $ J $ of independent information streams. However, this trend need not apply when measuring $ \dintdlj $ relative to the transmitted information $ \infoj $ or to $ \dinidlj $. Should it decrease instead, the destructive interference would become a minor component of $ \dinidl $, and according to current beliefs, would play a minor role in the cost of ignoring noise correlations in optimal decoding. In this section, we test this hypothesis and show that the relative average of $ \dintdlj $ never decreases, thereby driving $ \dinidl_\sj $ towards, but not necessarily reaching, its upper bound $ \dinid_\sj $.

\begin{figure}[htb!]
\centering
\includegraphics{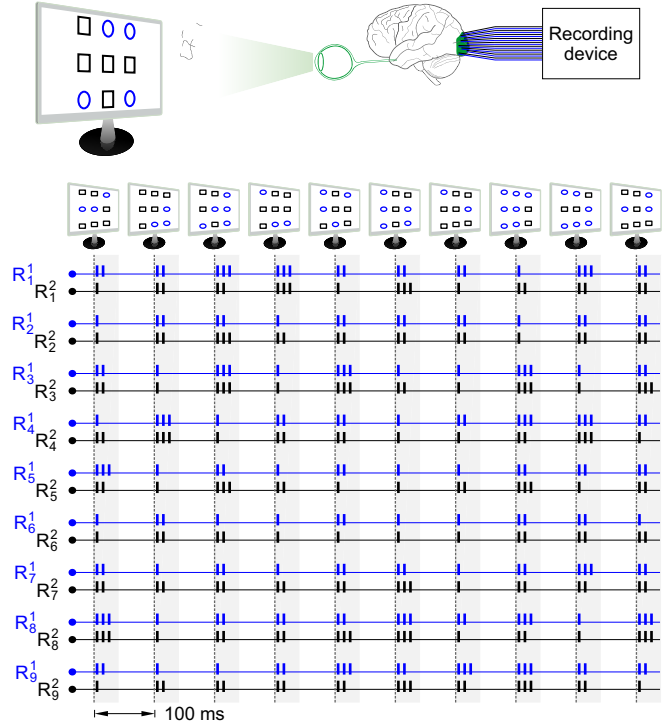}
\caption[]{Hypothetical experiment with $ 9 $ populations of two neurons each firing independently and selectively to $ 9 $ different and independently chosen stimulus features. Each feature is a frame ($ \SBox $ or $ \SCirc $) located at a different position in the screen. The populations were generated by replicating the first population in \fref{fig03}(a), except that $ P(S_j{=}\SBox $ and $ P(\VR_j{=}[2,2]|S_j{=}\SBox)$ were chosen independently for each population.}\label{fig06}
\end{figure}

To test this hypothesis, we build an experiment with $ J $ neural populations analogous to the first population in \fref{fig03}. These populations fire in response to visual stimuli $ \VS{=}[S_1,\ldots,S_J] $ composed of $ J $ different and independently-chosen stimulus features. Each $ S_j $ denotes a frame ($ \SBox $ or $ \SCirc $) projected at a different location in a screen. For each value of $ S_j $, the $ \jth $ population produces only two types of responses (\fref{fig06}).

In this experiment, $ P(\VS,\VR) $ was chosen to fulfill \eref{met::eq::indjoint}. In this way, each $ \jth $ population fires independently of the other populations and selectively to the $ \jth $ stimulus feature $ S_j $, thereby constituting independent information streams. For each stream, we set $P(\VR_j{=}[2,3]|S_j{=}\SCirc)$ equal to  $P(\VR_j{=}[2,2]|S_j{=}\SBox) $, and chose $ P(S_j{=}\SBox) $ and $ P(\VR_j{=}[2,2]|S_j{=}\SBox) $ both uniformly from the interval $ [0.05,0.95] $, and independently for each $ \jth $ population.

\begin{figure}[htb!]
\centering
\includegraphics{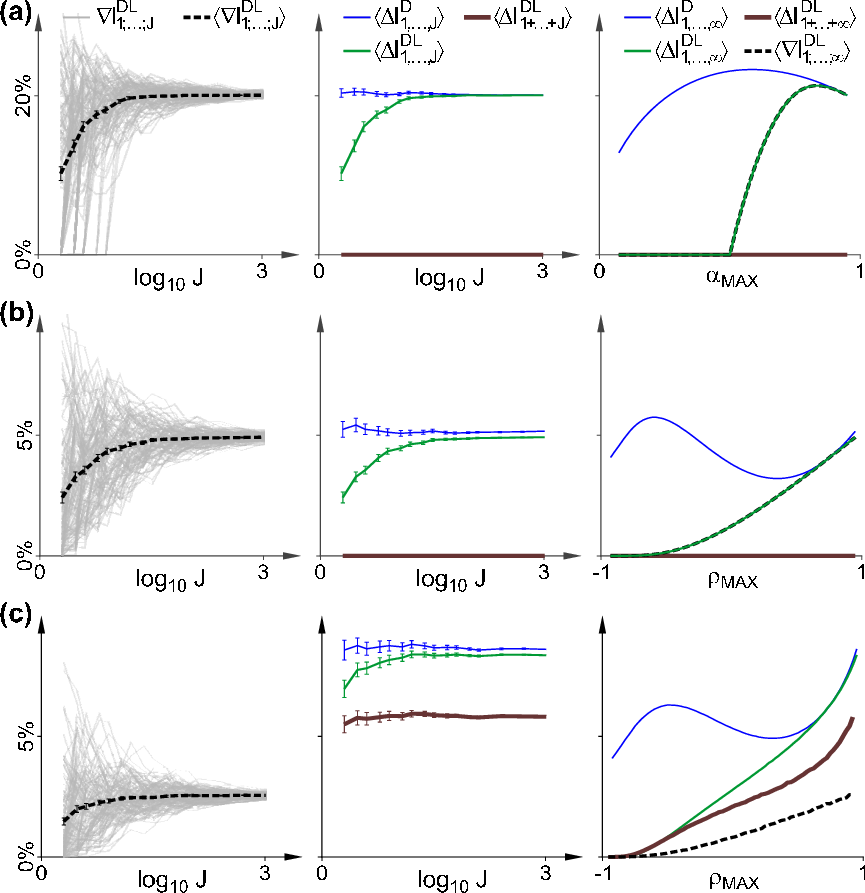}
\caption[]{Destructive interference grows with the number of independent-information streams. (a) Variation as the number $ J $ of streams grows, for 128 instances of hypothetical experiments generated from \fref{fig05}(a), of the destructive interference (left), its average across instances (left) and the averages of the information losses $\lrang{\dinid_\sj} $,$\lrang{ \dinidlpj} $ and $\lrang{ \dinidlj} $ (middle). The right panel shows the value attained by the destructive interference and the above information losses in the limit for large $ J $, when all $ P(\VR_j{=}[2,2]|S_j{=}\SBox) $ are drawn uniformly from the interval $[0.05,\alpha_{M\!A\!X}] $, as a function of $ \alpha_{M\!A\!X} $ (right panel). All values are given relative to $ \infoj $. (b) Analogous description to (a), but for 128 instances of hypothetical experiments generated from \fref{fig05}(b), and with correlation coefficients in the right panel drawn from the interval $[-0.95, \rho_{M\!A\!X}] $. (c) Analogous description to (b), but for 128 instances of hypothetical experiments generated from \fref{fig05}(c). Vertical segments: standard error of the mean.}\label{fig07}
\end{figure}

Repeating the above procedure, we generated 128 random instances of the same hypothetical experiment, each one with 1024 independent information streams. For each instance, we computed $ \dintdlj$ using the first $ J $ streams. This computation showed that the relative value of $ \dintdlj $ can increase or decrease with $ J $ depending on the sample (\fref{fig07}(a), left). However, we also found that the average value of $ \dintdlj $ across instances never decreased.

For the reasons we mentioned in \sref{res::sec::brain}, we repeated the analysis using as prototype information streams not the first population in \fref{fig03}, but the amplitudes of the $ 10\,Hz $-oscillations in the activity of two cortical areas studied in \fref{fig04}. The resulting experiments can be thought as concerning the information transmitted independently by the amplitudes of $ 10\,Hz $-oscillations at $ J $ pairs of cortical areas, namely

\begin{align}
Z_j^1 &= R_j^1\,\sin(20\,\pi\,t)\\
Z_j^2 &= R_j^2\,\sin(20\,\pi\,t)\, ,
\end{align}

\noindent where $ Z_j^k $ denotes the activity recorded in the $ \kth $ cortical area of the $ \jth $ pair. However, the experiments can also be interpreted as the amplitudes of oscillations within $ J $ different frequency bands $ f_1,\ldots,f_J $ in the activity of two cortical areas, namely

\begin{align}
Z_1 &= \sum_{j=1}^{J}{R_j^1\,\sin(2\,\pi\,f_j\,t)}\\
Z_2 &= \sum_{j=1}^{J}{R_j^1\,\sin(2\,\pi\,f_j\,t)}\, ,
\end{align}

\noindent or a combination of both.

As before, the stimulus-response probabilities were chosen to fulfill \eref{met::eq::indjoint}. Each 
$ P(\VR_j|S_j) $ was given by the following unit-variance Gaussian distribution
\begin{equation}
P(\VR_j|S_j) \propto \exp\lrcor{\frac{\big(R_j^1-R_j^2\big)^2}{2\,({\rho_j(S_j)}^2-1)}-\frac{(R_j^1-\mu_j)\,(R_j^2-\mu_j)}{1+\rho_j(S_j)}}\, ,
\end{equation}

\noindent where $ \rho_j(S_j) $ is the correlation coefficient between $ R_j^1 $ and $ R_j^2 $ given $ S_j $; whereas $ \mu_j {=} 2$ if $ S_j {=} \SBox $ and $ 4$ if $ S_j {=} \SCirc$. We chose $ P(S_j{=}\SBox) $ and all correlation coefficients independently for each stream, the former uniformly from the interval $ [0.05,0.95] $, and the latter uniformly from the interval $ [-0.95,0.95] $. Analogously to \fref{fig05}, we considered both the case in which the correlation coefficients $ \rho_j(R_j^1,R_j^2|\SBox) $ and $ \rho_j(R_j^1,R_j^2|\SCirc) $ for each $ \jth $ information stream coincide (\fref{fig07}(b)), and the case in which $ \rho_j(R_j^1,R_j^2|\SBox) $ and $ \rho_j(R_j^1,R_j^2|\SCirc) $ differ (\fref{fig07}(c)). In both cases, we found trends for the relative value and the relative average of $ \dintdlj $ as a function of $ J $ analogous to those in \fref{fig07}(a).

We also found that, for the three hypothetical experiments mentioned above, the relative averages of $ \dinid $ and $ \dinidlpj $ remain virtually constant as $ J $ grows (\fref{fig07}, middle panels). In that case, the growth with $ J $ of the average $ \dintdlj $ relative to $ \infoj $ is equivalent to that of the average $ \dinidlj $ relative to $ \infoj $. Most importantly, these results imply that the relative value of $ \dintdlj $ with respect to $ \dinidlj $ also grows with $ J $. In other words, the proportion of $ \dinidlj $, as opposed to that of $ \infoj $, explained by $ \dintdlj $, never decreases with the number of independent information streams.

The above results are valid not only for the three hypothetical experiments studied in \fref{fig07},  but for more general hypothetical experiments comprising $ J $ independent information streams defined through independent and identically distributed parameter vectors. For these class of experiments, we can prove that the following relations hold

\begin{align}
\lrang{I_{1,\ldots,\lambda J}} & = \lambda J\,\lrang{I_j}\label{met::eq::infolinear}\\
\lrang{\dinid_{1,\ldots,\lambda J}} & = \lambda J\,\lrang{\dinid_j}\label{met::eq::dinidlinear}\\
\lrang{\dinidl_{1\mtp\ldots\mtp\lambda J}} & = \lambda J\,\lrang{\dinidl_j}\label{met::eq::dinidlplinear}\\
\lrang{\dinidl_{1,\ldots,\lambda J}} &\geq \lambda \,\lrang{\dinidl_\sj}\label{met::eq::dinidlgreaterJ}\\
\lrang{\dintdl_{1;\ldots;\lambda J}} &\geq \lambda \,\lrang{\dintdlj}\label{met::eq::dintdlgreaterJ}\, ,
\end{align}

\noindent where $ \lambda $ is a positive integer. All these equations rest partially on the linearity of the mean. In addition, \eref{met::eq::infolinear} stems from \eref{met::eq::infoadd}; \eref{met::eq::dinidlinear}, from \eref{res::eq::dinidlt12} with $ \theta{=}1 $; \eref{met::eq::dinidlgreaterJ}, from the convexity of $ \dinitdl $; and \eref{met::eq::dintdlgreaterJ}, from subtracting \eref{met::eq::dinidlgreaterJ} and \eref{met::eq::dinidlplinear}.

Dividing by $ \lrang{I_{1,\ldots,\lambda J}} $ each side of \eref{met::eq::dinidlinear}--\eref{met::eq::dintdlgreaterJ} yields the following relations for the relative averages 

\begin{align}
\lrang{\dinid_{1,\ldots,\lambda J}}/\lrang{I_{1,\ldots,\lambda J}} & = \lrang{\dinid_j}/\lrang{I_j}\label{met::eq::dinidrelconstant}\\
\lrang{\dinidl_{1\mtp\ldots\mtp\lambda J}}/\lrang{I_{1,\ldots,\lambda J}} & = \lrang{\dinidl_j}/\lrang{I_j}\label{met::eq::dinidlprelconstant}\\
\lrang{\dinidl_{1,\ldots,\lambda J}}/\lrang{I_{1,\ldots,\lambda J}}  &\geq \lrang{\dinidl_\sj}/\lrang{I_\sj}\label{met::eq::dinidlrelgreaterJ}\\
\lrang{\dintdl_{1;\ldots;\lambda J}}/\lrang{I_{1,\ldots,\lambda J}}  &\geq \lrang{\dintdl_\sj}/\lrang{I_\sj}\label{met::eq::dintdlrelgreaterJ}\, .
\end{align}

\noindent Furthermore, dividing \eref{met::eq::dintdlgreaterJ} by $ \lrang{\dinidl_{1,\ldots,\lambda J}} $ yields the following relation

\begin{equation}
\lrang{\dintdl_{1;\ldots;\lambda_1 \lambda_2 J}}/\lrang{\dinidl_{1,\ldots,\lambda_1 \lambda_2 J}}  \geq \lrang{\dintdl_{1;\ldots;\lambda_1 Jj}}/\lrang{\dinidl_{1,\ldots,\lambda_2 J}}\label{met::eq::dintdlreldinidlgreaterJ}\, ,
\end{equation}

\noindent after replacing $ \lambda $ with the product of two positive integers $ \lambda_1 $ and $ \lambda_2 $, and some relatively simple algebra. These relations generalize our observations in \fref{fig07}, and prove that the destructive interference never decreases in absolute or relative magnitude, and becomes increasingly important as the number of independent information streams grows.

We also noticed that the relative magnitude of $ \dintdlj $ converges as $ J $ grows to a value, here called $ \dintdl_\sintinf $. This observation need not be surprising for the relative average of $ \dintdl_\sintinf $ across instances. Indeed, not only this value is bounded by unity, but also increasing with $ J $, as we have shown, and therefore the monotone convergence theorem ensures that a limit exists as $ J{\rightarrow}\infty $. However, our observation also applies to each instance of the experiments, as opposed to their averages (\fref{fig07}, left). Most importantly, these observations open up the possibility that $ \dintdlj $ becomes so large when $ J{\rightarrow}\infty $ as to drive $ \dinidlj $ close to its maximum value, namely $ \dinid_\sj $.

To test this hypothesis, we generated different instances of the experiment in \fref{fig07}(a) by choosing $ P(\VR_j{=}[2,2]|S_j{=}\SBox) $ uniformly from intervals $ [0.05,\alpha_{M\!A\!X}]$, for $ 0.05{\leq}\alpha_{M\!A\!X}{\leq}0.95 $. For these instances, we computed $ \info_\sdinf $ analytically, and estimated $ \dinid_\sdinf $, $ \dinidl_\sdinf $ and $ \dinidl_\spinf $ numerically using Matlab R2015b.

We found that the relative value of $ \dintdl_\sintinf $, and consequently of $ \dinidl_\sdinf $, reached its maximum when $\alpha_{M\!A\!X}{\approx}0.83 $. This value was estimated as the maximum of a cubic function that locally approximated the trace of $ \dint_{1,\ldots,\infty} $ as $ \alpha_{M\!A\!X} $ grows. Furthermore, we found that both $ \dintdl_\sintinf $ and $ \dinidl_\sdinf $ converged to $ \dinid_\sdinf $ whenever $ P(\VR_j{=}[2,2]|S_j{=}\SBox) $ was drawn from intervals centered at $ 0.5 $. Indeed, for large $ J $, $ \dinidltsub{}{\sj} $ can be approximated using the law of large numbers as follows

\begin{equation}
\dinidltsub{}{\sj} \approx J\,\lrang{q_j\,\alpha_j\,\ln\left(1+\kqj\kalphajt\right)} \, ,
\end{equation}

\noindent where for compactness, we have employed a notation analogous to that introduced after \eref{res::eq::dinidl12fig03}, defining $ q_j{=}P(S_j{=}\SBox) $, $ \alpha_j{=}P(\VR_j{=}[2,2]|S_j{=}\SBox) $, and $ \bar{x}{=}1{-}x $ for any real value $ x $. After some algebra, this equation can be shown to reach its minimum when $ \theta{=}1 $ provided that $ \qj $ and $ \alphaj $ are chosen with probability distributions that are symmetric about $ 0.5 $, and hence

\begin{equation}
\dinidlj \llnapprox \dinid_\sj \, . 
\end{equation}

In \fref{fig07}(b)-(c), different instances were generated by choosing the correlation coefficients uniformly from intervals $ [-0.95,\rho_{M\!A\!X}]$, for $ -0.95{\leq}\rho_{M\!A\!X}{\leq}0.95 $. In these cases, we estimated $ I_\sdinf $, $ \dinid_\sdinf $ and $ \dinidl_\sdinf $ using Monte Carlo integration in Python 3.4.3 with the packages Vegas 3.0 and SciPy 0.14.1, using 2000000 vector samples, divided in 10 iterations for training and 10 iterations for evaluation. The value of $ \dinidl_\spinf $ was estimated using $ 163840 $ independently generated streams.

Unlike in \fref{fig07}(a), here the relative values of both $ \dintdl_\sintinf $ and $ \dinidl_\sdinf $ increased with $ \rho_{M\!A\!X} $. However, $ \dinidl_\sdinf $ converged to $ \dinid_\sdinf $ only when $\rho_{M\!A\!X}{\approx}0.73 $ in both \fref{fig07}(b) and \fref{fig07}(c). This value was estimated analogously to the value of $ \alpha_{M\!A\!X} $ in \fref{fig07}(a).

To summarize, our results show that as the number of independent information streams grows, the relative amount of destructive interference is never negative and never decreases. This result finally answers the question we posited in the previous section, and shows that $ \dinidl $ may grow with the number of neurons or the decoding-window duration by the virtue of destructive interference alone, or in conjunction with other possible causes. In addition, destructive interference may drive the loss $ \dinidlj $ near its upper bound $ \dinid_\sj $. Analogous observations have previously been regarded as a sign that $ \dinid_\sj $ is tight \citep{latham2013}. Our results once again cast doubt on this interpretation, but this time by showing that it may stem from a paradoxical growth in the cost of ignoring noise correlations.

\subsection{The construction of optimal NI decoders is not unique}\label{res::sec::moreefficient}

The measure $ \dinidlj $ computed above stems from computing the measure $ \dinidl $ introduced by \citet{latham2005} using all neurons in $ J $ populations that transmit independent information. According to previous studies, this computation would yield the exact information loss caused when optimally decoding the aforementioned neurons ignoring noise correlations. However, in this section we show that $ \dinidl $ is actually limited to a specific construction of optimal NI decoders, thereby opening up the possibility that other constructions be more efficient than predicted by $ \dinidl $.

\begin{figure}[htb!]
\centering
\includegraphics{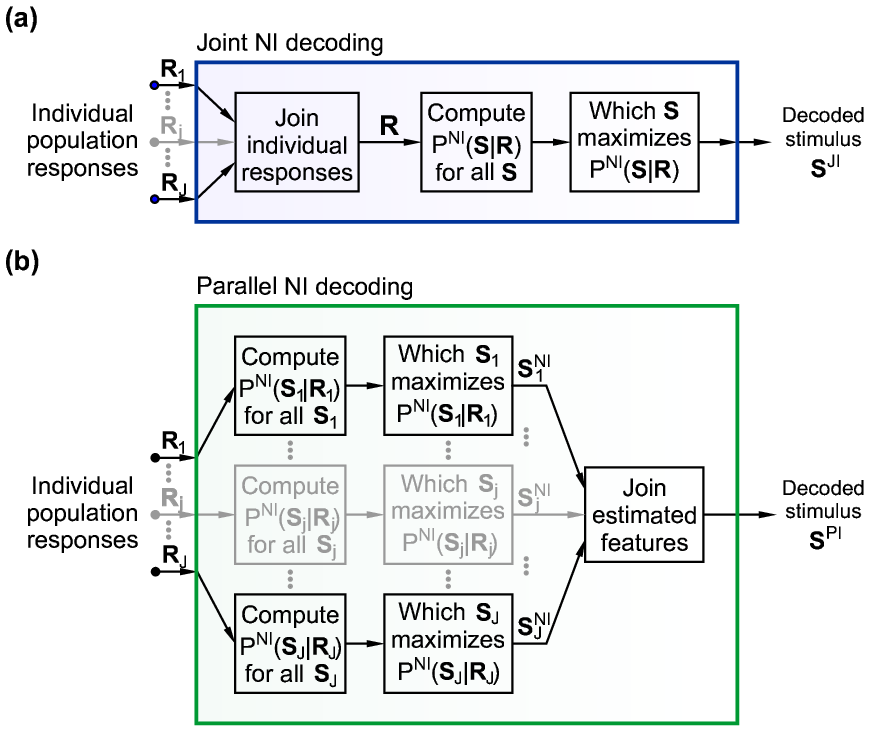}
\caption[]{Two different constructions of optimal NI decoders. (a) Joint NI decoders choose stimultaneously all the stimulus features that maximize the NI posterior $ \PNIs(\VS|\VR) $. (b) Parallel NI decoders choose each stimulus feature singly but in parallel by maximizing each of their corresponding NI posteriors $ \PNIs(S_j|\VR_j) $.}\label{fig5}
\end{figure}

Specifically, we noticed that $ \dinidl $, and consequently $ \dinidlj $, use what we here call joint NI decoders \citep[also called centralized integration;][]{zhang2016}. These decoders produce simultaneous estimates $ \SJI $ of all stimulus features only after reading the concurrent responses $ \VR $ of all populations (\fref{fig5}(a)). Mathematically, joint NI decoders can be defined as follows

\begin{equation}\label{met::eq::sji}
\SJI = \arg \max_{\VS}{\PNIs(\VS|\VR)}\, .
\end{equation}

\noindent However, joint NI decoders need not be the only way in which optimal NI decoders can be constructed.

To show this, recall our definition of independent information given in \sref{met::sec::indinfo}. According to this definition, $ J $ neural populations transmit independent information when they fire independently and selectively to $ J $ independent stimulus features, respectively. In that case, the stimulus features can also be optimally identified in parallel \citep[also known as distributed, modular or decentralized decoding;][]{zhang2016} regardless of whether neurons are noise independent or not. However, the use of parallel decoders when studying the role of noise correlations in neural decoding has previously been controversial \citep{meister2001,nirenberg2001,schneidman2003,latham2005}.

As we note here, this controversy could have been avoided should previous studies have combined the outputs of the parallel decoders into a single consistent estimate. However, such combination may not be possible without taking implicitly or explicitly correlations into account or making additional assumptions \citep{landy1995,knill1996,schneidman2003,jaynes2003,eyherabide2013}. These difficulties do not arise here because populations transmit independent information.

Specifically, here we construct parallel NI decoders as two-stage processes (\fref{fig5}(b)). The first stage consists of $ J $ optimal NI decoders, each of which operates separately on a different population. Mathematically, the $ \jth $ optimal NI decoder reads the response $ \VR_j $ of the $ \jth $ population and produces an estimate $ S^{PI}_j$ of the $ \jth $ feature, according to the following

\begin{equation}\label{met::eq::spi}
S^{PI}_j = \arg \max_{S_j} \PNIs(S_j|\VR_j) = \arg \max_{S_j} \PNIs(S_j|\VR)  \, .
\end{equation}

\noindent These estimates are concurrently fed into the second stage, which concatenates them to produce the estimated stimulus $ \SPI{=}{\lrcor{S^{PI}_1,\ldots,S^{PI}_J}} $. Unlike \citet{zhang2016}, the definitions of joint and parallel decoders given here are both valid for arbitrary stimulus distributions.

The last equality in \eref{met::eq::spi} shows that feeding each $ \jth $ optimal NI decoder in the first stage only with the responses of the $ \jth $ population (or demultiplexing, as mentioned in \sref{res::sec::brain}) is inessential. Indeed, after some algebra, \eref{se::eq::infoaddtransform} turns into $ \PNI(S_j|\VR_j){=}\PNI(S_j|\VR) $ for populations that transmit independent information. Therefore, parallel estimations of each individual stimulus feature $ S^{PI}_j $ can be conducted without the interference of other information streams that seemingly affect the joint estimations, at least according to the results we have obtained above.

To summarize, we have shown for the first time that $ \dinidl $ is only based on joint NI decoders. However, even though unnoticed by previous studies, we found that the construction of optimal NI decoders need not be unique, and different alternatives may exist depending on the case. In particular, we showed that, when information is independent, optimal NI decoding can be performed in parallel. These possibilities have remained largely ignored in the neural coding literature, and most importantly, may lead to constructions that outperform joint NI decoders and overcome the destructive interference. In that case, $ \dinidl $ would potentially overestimate $ \Delta CI $, and the importance of noise correlations in optimal decoding, for yet another reason than those we have mentioned above.

\subsection{Decoding more efficiently than predicted}

Our finding that the construction of optimal NI decoders is not uniquely defined immediately raises the question of what difference do different constructions make. To address this question, we compute in this section the information losses caused by parallel NI decoders using the information notions underlying $ \dinid $, $ \dinidd $ and $ \dinidl $. Contrary to current pervasive ideas, these computations need not stem straight-forwardly from previous results on encoded information losses for for the reasons we mentioned in \sref{res::sec::difnotion}.

The axiomatic information loss caused by parallel NI decoders can be computed by reinterpreting the first stage as parallel transformations of each individual population response. When information is independent, \eref{se::eq::infoaddtransform} holds, and the axiomatic information losses $ \dinidl_1,\ldots,\dinidl_J $ caused by each of these parallel transformations are additive. The second stage is invertible, and hence lossless. Consequently, using the notation introduced in \sref{met::sec::indinfo}, the axiomatic information loss caused by parallel NI decoders is equal to $ \dinidd_\spj $.

Analogous results can be obtained for the value of $ \dinid $ associated with parallel NI decoders. To that end, recall that $ \dinid $ was derived by \citet{nirenberg2001} using a notion of information with roots in coding theory, here called descriptive information. Within this notion, $ \dinid $ can be interpreted as the increment caused by ignoring noise correlations in the minimum average description length $ L_{\VS} $ of the stimulus identity $ \VS $ after observing the population response $ \VR $ \citep{cover2006}.

Within that context, parallel NI decoding can be interpreted as describing $ \VS $ by concatenating the individual descriptions of its $ J $ stimulus features. The resulting minimum average description length $ L^{PI}_{\VS} $ equals the sum of the minimum average description lengths $ L_{S_1},\ldots,L_{S_J} $ associated with each stimulus feature, and therefore, their corresponding increments $ \dinid_1,\ldots,\dinid_J $ caused by ignoring noise correlations are additive. Consequently, the descriptive information loss caused by parallel NI decoders equals $ \dinid_\spj $. Notice that this result need not immediately arise from the additivity of $ \dinid $ found in \sref{res::sec::ubiquity}, for that mathematical findings need not be conceptually related to the operation of parallel NI decoders.

The above results render seemingly reasonable to hypothesize that the communication information loss produced by parallel NI decoders equals $ \dinidlpj $. To rigorously prove this, recall the notation introduced in \sref{met::sec::indinfo}, and the derivation of \citet{latham2005} introduced in \sref{res::sec::difnotion}. According to their derivation, we can associate each $ \jth $ population with codebooks of up to $ \exp(N\,\tilde{I}_j)$ sequences $ \MSEnc[j]$, where $ \tilde{I}_j{\leq} I_j{-}\dinidl_j $, for which the average decoding-error rate $ \PSNIerrj $ produced by optimal NI decoders vanishes exponentially as $ N $ grows.

Combining the above codebooks using Cartesian products yields a product codebook of up to $ \exp(N\,\sum_{j=1}^{J}{\tilde{I}_j})$ sequences. For theses codebooks, Boole's inequality \citep{casella2002} yields after some algebra that the 
average decoding error rate $ \PSNIerr $ decays exponentially as $ N $ grows at least as fast as $ \max_j{\PSNIerrj} $. Hence, $ \dinidlpj $ is achievable \citep{cover2006} and quantifies the communication information loss caused when decoding $ J $ independent information streams ignoring noise correlations.

In conclusion, we have proved that, when information is independent, the information loss caused by parallel NI decoders is equal to the sum of the information losses caused by each of its constituent optimal NI decoders regardless of the underlying information notion. Most importantly, this result shows that $ \dinidlpj $ is achievable, thereby proving that parallel NI decoders can overcome the destructive interference. Therefore, we conclude that parallel NI decoders can be more efficient than predicted by $ \dinidlj $, and that, contrary to current beliefs, noise correlations can be less important than predicted by $ \dinidl $.

\subsection{Joint NI decoders can potentially outperform parallel NI decoders}

We have shown in the previous sections have shown that the communication information loss $ \dinidlpj $ caused by parallel NI decoders is never greater and can be less than the communication information loss $ \dinidlj $ caused by joint NI decoders. These results seemingly indicate both that parallel NI decoders do outperform joint NI decoders, at least in terms of communication information losses, and that, regardless of how paradoxical it may seem, destructive interference need not merely stems from flaws in $ \dinidl $. In this section, we address these hypotheses and show that joint NI decoders can both achieve $ \dinidlpj $ and potentially outperform parallel NI decoders, at least in terms of axiomatic information loss.

To prove that joint NI decoders can achieve $ \dinidlpj $, recall that the derivation of $ \dinidl $ in \citet{latham2005} is based on the average decoding error probability $ \PSNIerr $, which they wrote as $ P\left(\PNIs(\VS|\VR){<}\PNIs(\VSNI|\VR)\right) $. However, notice that the estimates $ \SJI $ and $ \SPI $ produced by joint and parallel NI decoders, respectively, are always associated with the same NI posteriors, namely

\begin{align}
\PNIs(\SJI|\VR) &=\max_{\SSE_1,\ldots,\SSE_J}{\textstyle\prod_{j=1}^J{\PNI(S_j|\VR_j)}}\nonumber\\
 &={\textstyle\prod_{j=1}^J}{\max_{S_j}{\PNIs(S_j|\VR_j)}}\nonumber\\
&=\PNIs(\SPI|\VR)\, . \label{res::eq::equalest}
\end{align}

\noindent Therefore, $ P\left(\PNIs(\VS|\VR){<}\PNIs(\SJI|\VR)\right){=} P\left(\PNIs(\VS|\VR){<}\PNIs(\SPI|\VR)\right) $, and consequently joint and parallel NI decoders produce the same average decoding error probability for the same set of codebooks. In other words, joint NI decoders can also achieve $ \dinidlpj $, thereby rendering $ \dinidlj $ as an upper bound of the actual communication information loss caused by joint NI decoders.

The equality between $ \PNIs(\SJI|\VR) $ and $ \PNIs(\SPI|\VR) $ also implies that both decoders will typically produce the same estimates. As a result, the axiomatic information losses $ \dinidd_\sj $ and $ \dinidd_\spj $ caused by joint and parallel NI decoders, respectively, will typically coincide. However, differences in their estimates, and hence in their axiomatic information losses, can still stem from arbitrary tie-braking rules.

Specifically, the maxima in \eref{met::eq::sji} and \eref{met::eq::spi}, respectively, oftentimes occur for multiple stimuli. Hence, joint and parallel NI decoders cannot unambiguously produce an estimate. In practice, these situations need not be rare and may arise, for example, due to probability quantization when estimating them from experimental frequencies \citep{samengo2002,casella2002}. Should ambiguities arise, they can be resolved by adopting tie-breaking rules. Depending on how they are set, $ \dinidd_\sj $ can be greater, equal or less than $ \dinidd_\spj $.

\begin{figure}[htb!]
\centering
\includegraphics{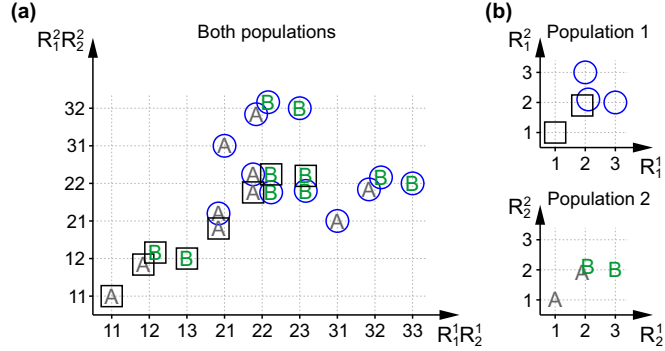}
\caption[]{Experiment in which joint and parallel NI decoders can yield different information losses even though information is independent. (a) Analogous description to \fref{fig03}(a). (b) Analogous description of \fref{fig03}(b). 
}\label{fig7}
\end{figure}

To illustrate this, we build the hypothetical experiment shown in \fref{fig7}(a). Analogously to \fref{fig03}(a), this experiment consists of two populations that fire independently and selectively to independent stimulus features. However, here population 1 also produces the response $ \VR_1{=}[2,2] $ after $ \SCirc $ (where $ \VR_j{=}[R_j^1,R_j^2] $). In addition, population 2 includes a second neuron that produces the same number of spikes as the first neuron after $ \SA $, and only two spikes regardless of the response of the first neuron after $ \SB $ (\fref{fig7}(b)).

Suppose that we set $ P(S_1,\VR_1)$ so that $ P(\SBox,[2,2]){=}12/30$, $P(\SCirc,[2,2]){=}7/30$ and $P(\SCirc,[2,3]){=}2/30 $. In that case, $ \PNIs(\SBox|\VR_1){=}\PNIs(\SCirc|\VR_1) $ only when $ \VR_1{=}̛[2,2] $, and thus optimal NI decoders cannot unambiguously choose a frame. To resolve this ambiguity, we can adopt the convention of choosing $ \SCirc $ over $ \SBox $, here denoted $ \SCirc{>}\SBox $. This convention minimizes the axiomatic information loss $ \dinidd_1 $ computed using only population 1 ($ {\approx} 31\,\% $).

Analogously, suppose that we set $ P(S_2,\VR_2)$ so that $ P(\SA,[2,2]){=}90/190$ and $P(\SB,[2,2]){=}81/190$. In that case, $ \PNIs(\SA|\VR_2){=}\PNIs(\SB|\VR_2) $ only when $ \VR_2{=}̛[2,2] $, and thus optimal NI decoders cannot unambiguously choose a letter. To resolve this ambiguity, we can adopt the convention $ \SA{>}\SB $, which minimizes the axiomatic information loss $ \dinidd_2 $ computed using only population 2 ($ {\approx}11\,\%$).

Together, the above two conventions minimize the axiomatic information loss $ \dinidd_{1+2} $ caused by parallel NI decoders (${\approx}42\,\%$). On the contrary, the opposite conventions (i.e., $ \SCirc{<}\SBox $ and $ \SA{<}\SB $) maximize $ \dinidd_{1+2} $ (${\approx }46\,\% $; with $ \dinidd_1{\approx}34\,\%$ and $ \dinidd_2{\approx}12\,\%$). Other two ways of combining the conventions exist that yield intermediate values of $ \dinidd_{1+2} $, thereby adding up to four different constructions of parallel NI decoders.

On the contrary, the number of joint NI decoders that can be constructed by choosing different tie-breaking rules is 64. These rules must choose between 

\begin{enumerate}
\item $ \SBA $ and $ \SCA $ if $ \VR{=}[2,2,1,1] $, because $ \PNIs(\SBA|\VR){=}\PNIs(\SCA|\VR){=}0.5$; 
\item $ \SBA $ and $ \SBB $ if $ \VR{=}[1,1,2,2] $, because $ \PNIs(\SBA|\VR){=}\PNIs(\SBB|\VR){=}0.5$; 
\item $ \SCA $ and $ \SCB $ if $ \VR{=}[2,3,2,2] $ or $ [3,2,2,2] $, because $ \PNIs(\SCA|\VR){=}\PNIs(\SCB|\VR){=}0.5$; 
\item $ \SBB $ and $ \SCB $ if $ \VR{=}[2,2,2,3] $, because $ \PNIs(\SBB|\VR){=}\PNIs(\SCB|\VR){=}0.5$; and 
\item all stimuli if $ \VR{=}[2,2,2,2] $, because $ \PNIs(\VS|\VR){=}0.25 $ for all $ \VS $.
\end{enumerate}

\noindent By adopting tie-breaking rules independently for each of the above five situations, we can find joint NI decoders that cause axiomatic information losses $ \dinidd_{1,2} $ as low as $ {\approx}16\,\% $ (only $ {\approx}38\,\% $ of the minimum $ \dinidd_{1+2} $) or as large as $ {\approx} 58\,\% $ (and thus $ {\approx} 28\,\% $ larger than the maximum $ \dinidd_{1+2} $).

In conclusion, we have shown that joint and parallel NI decoders are almost always equivalent, not only in terms of axiomatic information losses, namely $ \dinidd_\sj{=}\dinidd_\spj $, but also in their stimulus estimates, namely $ \SJI $ and $ \SPI $. However, this need not be the case when tie-breaking rules must be chosen. In those cases, $ \dinidd_\spj$ can be greater or less than  $\dinidd_\sj $ depending on the chosen conventions, but, without restrictions, the best parallel NI decoder can never outperform the best joint NI decoder.

These results may also apply to communication information losses $ \dinidl_\spj $ and $ \dinidl_\sj $. Indeed, we have shown that both joint and parallel NI decodes can achieve the same communication information losses, namely $ \dinidlpj $. However, this need not always be the case for at least three reasons. First, neither our computations nor the derivation of $ \dinidl $ in \citet{latham2005} have taken into account the potential effect of tie-breaking rules. Second, here we showed that the set of all codebooks employed by \citet{latham2005} to derive $ \dinidl $ may produce larger average decoding error probabilities than some of its subsets. Third, the set of codebooks for which joint and parallel NI decoders achieve the minimum average decoding error probability, respectively, need not coincide.

These three considerations can be proved unnecessary when studying other response aspects such as spike counts or latencies. However, for the reasons we mentioned in \sref{res::sec::difnotion}, this observation need not apply to the study of noise correlations from the decoding perspective. 
Consequently, our results open up the possibility that joint NI decoders outperform parallel NI decoders in terms of communication information loss. Most importantly, they allow us to finally refine the estimation of $ \Delta CI $ as follows

\begin{equation}
\Delta CI \begin{cases}=0 & \mbox{if $ P(\VS{\neq}\VSD){=}0 $}\\
\leq\dinidlpj & \mbox{if information is independent}\\ 
\leq\dinidl & \mbox{otherwise} \end{cases}\, .
\end{equation} 

\noindent Above all, our results in this section show that the overestimation of $ \dinidl $ need not be constrained to independent information streams, and that current beliefs in the exactness of $ \dinidl $ may lead one to overestimate the importance of noise correlations in optimal decoding.

\section{Discussion}

Many measures have been proposed to quantify the information loss caused by ignoring noise correlations in neural decoding, but their conceptual and quantitative accuracy remains controversial \citep{nirenberg2001,meister2001,nirenberg2003,schneidman2003,latham2005,averbeck2006,oizumi2009,ince2010,oizumi2010,eyherabide2013,latham2013}. Resolving these controversies is fundamental for understanding the role of noise correlations in brain computations over multiple groups of neurons or neural substrates, or when decoding brain signals recorded from multiple brain locations either in one or even multiple subjects \citep{hari2009,babiloni2014}. In this study, we focus on one of the most prominent measures, due to its information-theoretical foundations and its underlying communication notion of information, here called $ \dinidl $. This measure was introduced by \citet{latham2005} based on the work of \citet{merhav1994} on mismatched decoding, and it is currently considered the exact information loss caused by ignoring noise correlations in optimal decoding \citep{latham2005,oizumi2009,ince2010,oizumi2010,latham2013,oizumi2016}, despite the fact that, to our knowledge, the properties of $ \dinidl $ and the consequences of its putative exactness have both remained largely unexplored.

Accordingly, our first step in this direction was to address the implications of what to our knowledge is the only controversial finding to date concerning $ \dinidl $ and $ \dinidd $. As a putative exact measure of information loss, $ \dinidl $ would be expected to never exceed $ \dinidd $ \citep{cover2006,quiroga2009,eyherabide2010b}. However, as we have recently shown, this need not be the case \citep{eyherabide2013}. However, due to the rigorous derivation of $ \dinidl $, it remains unclear whether this observation indicates major departures from traditional relations between information and decoding or flaws in $ \dinidl $.

Although puzzling, here we argued that these observations need not actually contradict the belief that $ \dinidl $ is exact for at least two reasons. First, as mentioned in \sref{res::sec::difnotion}, $ \dinidl $ and $ \dinidd $ are based on different notions of information, which differences are not necessarily unknown, but previous studies have often overlooked \citep{latham2005,thomson2005,quiroga2009,ince2010,oizumi2010,latham2013}. Second, here we point out that even though the relation $ \dinidl{<}\dinidd $ holds when studying response aspects such as spike counts or latencies \citep{eyherabide2016b}, this observation need not immediately imply that the relation must also hold when studying noise correlations.

These two reasons led us to hypothesize that previous observations of $ \dinidl$ exceeding $\dinidd $ may simply stem from fundamental differences between the two underlying information notions rather than from previously unforeseen flaws in $ \dinidl $. To disentangle these two possibilities, we compared for the first time the value of $ \dinidl $ with a direct computation of the communication information loss caused by ignoring noise correlations in optimal decoding that we called $ \Delta CI $. Contrary to currently thought \citep{latham2005,oizumi2009,ince2010,oizumi2010,latham2013,oizumi2016}, our results showed for the first time that $ \dinidl $ need not be exact and can overestimate $ \Delta CI $ at least when optimal NI decoders can perfectly identified some stimulus feature.

Using populations that transmit independent information, we also showed for that the first time that, that $ \dinidl $ is actually superadditive. Specifically, we found that the value of $ \dinidl $ computed using all neurons in all populations is larger than the sum of the values of $ \dinidl $ computed using all neurons in each population, respectively. This result constitutes a major departure from the traditional additivity of mutual information \citep{shannon1949,fano1961,cover2006}, thereby questioning current beliefs in the exactness of $ \dinidl $ \citep{latham2005,oizumi2009,ince2010,oizumi2010,latham2013,oizumi2016}.

This paradoxical increment in $ \dinidl $ was here called destructive interference, and shown ubiquitous regardless of whether the populations were interpreted as spatially or temporally multiplexed independent-information streams \citep{oppenheim1997,panzeri2010}, which we explained by tracing them back to the convex minimization in the definition of $ \dinidl $ \citep{latham2005,oizumi2009,oizumi2010,eyherabide2013,oizumi2016}. Our explanation implies that may arise regardless of the correlation importance within populations, the composition of neural substrates and the type of multiplexed codes. Furthermore, the mathematical nature of this result extends this phenomenon beyond neural coding to information streams of arbitrary type \citep{shannon1949,fano1961,ernst2002,eyherabide2010FCN,panzeri2010,oizumi2016,zhang2016}.

We also found that the destructive interference grows with the number of populations, driving $ \dinidl $ towards $ \dinid $ (\fref{fig07}), and possibly reaching ${\approx} 100\,\% $ of the transmitted information (\fref{fig03}), even when each population could be safely decoded ignoring noise correlations (figures~\ref{fig03};~\ref{fig04};~\ref{fig05};~\ref{fig07}). These results are qualitatively similar to previous experimental findings in which $ \dinidl $ grows with the number of neurons or with the decoding-window length, and lies close to $ \dinid $ \citep{oizumi2009,oizumi2010,latham2013}. Therefore, here we conclude that these phenomena may occur even in the absence of putative temporal correlations across time bins, pseudo-correlations caused by inappropriately assuming stationarity, or higher-order correlations, as previous studies have conjectured \citep{oizumi2009,oizumi2010}, due to the sole presence of destructive interference.

However, we found the emergence of destructive interference puzzling for at least two reasons. First, it never occurs when studying response aspects such as spike counts or latencies \citep{eyherabide2016b}, or when studying correlation importance using $ \dinid $, as we have here shown (\sref{res::sec::ubiquity}). Second, should $ \dinidl $ be exact as currently believed, noise correlations would seemingly grow in importance with the number of populations, even if they transmit independent information.

These reasons notwithstanding, we cannot rigorously conclude that they imply flaws in $ \dinidl $ for at least two other reasons. First, whether or not ignoring response aspects differs from ignoring response probabilities remains unsettled \citep{nirenberg2003,schneidman2003,eyherabide2013,latham2013}. Second, although often overlooked in previous studies \citep{latham2005,oizumi2009,oizumi2010,eyherabide2013,latham2013}, $ \dinid $ is based on a different notion of information than $ \dinidl $ \citep{nirenberg2001,nirenberg2003,latham2013} and may overestimate the communication information loss \citep{latham2005,oizumi2009,oizumi2010,eyherabide2013}. Instead, we hypothesized that the phenomenon of destructive interference may only indicate that the intuition gained from traditional information theory applied to ignoring aspects of neural responses should be observed with caution when applied to ignoring aspects of response probabilities.

Unnoticed by previous studies is the fact that $ \dinidl $ is based on only one possible construction of optimal NI decoders, that we called joint NI decoding, which identifies all stimulus features simultaneously (\fref{fig5}(a)). However, here we showed that optimal NI decoders can be constructed in other ways that can potentially outperform the state of the art. In particular, and despite previous controversies, here we proved that optimal NI decoders constructed to identify independent stimulus features separately but in parallel can completely overcome the destructive interference (\fref{fig5}(b)).

This finding was puzzling because, when information is independent, joint and parallel NI decoders typically produce the same estimates. This observation seemingly rules out construction differences as the ultimate cause of destructive interference, but provided us with valuable insight into potential flaws of $ \dinidl $, even when information is not independent. Specifically, we hypothesized that parallel NI decoders seemingly outperform joint NI decoders because they seemingly require that $ \dinidl $ be computed with codebooks that preserve the independence of the stimulus features.

Our tests on this hypothesis revealed for the first time in neural coding that, contrary to what occurs when studying response aspects such as spike counts or latencies, the average decoding error probability over the set of all codebooks used in the original derivation of $ \dinidl $ need not be representative of the average decoding error probability for smaller sets of codebooks when studying noise correlations. This result rigorously proved for the first time that, contrary to previously thought \citep{latham2005,oizumi2009,ince2010,oizumi2010,latham2013,oizumi2016}, $ \dinidl $ may overestimate the communication information loss caused by ignoring noise correlations in optimal decoding even when stimuli cannot be perfectly identified or when information is not independent. Most importantly, this observation puts forward
 testing different sets of codebooks as one possible strategy for solving the overestimation.

We point out that the concept of destructive interference is not limited to the communication notion of information, can also be extended to the axiomatic notion of information. From our results on joint and parallel decoding, it follows immediately that this axiomatic destructive interference is not related to overestimations of the axiomatic information loss but to differences in tie-breaking rules adopted during the construction of joint and parallel NI decoders. Furthermore, it can be positive or negative, but the minimum axiomatic destructive interference over all possible conventions is always negative. Whether tie-breaking rules play any role in the communication destructive interference, or whether joint NI decoders can actually extract more communication information than parallel NI decoders, still remain open questions.

Since its introduction, the information-theoretical measure $ \dinidl $
has been deemed the exact information loss caused by ignoring noise correlations in optimal decoding \citep{latham2005,oizumi2009,ince2010,oizumi2010,latham2013,oizumi2016}. However, our results prove that $ \dinidl $ is biased and that the overestimation can reach $ {\approx}100\,\% $ of the encoded information. Hence, using $ \dinidl $ in its basic form may lead to wasting experimental and computational resources, which can be avoided by estimating the communication information loss as we propose here. These estimates close the gap between axiomatic and communication information losses, thereby opening up the possibility that traditional relations between information and decoding observed when studying response aspects such as spike counts and latencies are also valid when studying noise correlations. In practice, our results indicate that noise correlations need not be as necessary as previously thought, and may potentially contribute to reduce the cost and complexity of computational brain models and neuroprosthetics.

\section{Conclusion}

Assessing the role of noise correlations in neural decoding is fundamental, not only for understanding how the brain perform computations and turn them into perceptions, decisions and actions, but also for estimating the amount of resources and the level of complexity required to study brain function and to construct neural prosthetics. This study sheds new light into their role by revealing and resolving unforeseen limitations of an approach that, due to its rigorous information-theoretical foundations, has always been deemed exact. Our analysis was entirely conducted taking into account the fundamental differences between the notions of information associated with this and other approaches. In this way, we avoided the confounds of previous studies, and rigorously proved that the currently-deemed-exact approach overestimates the information loss caused by ignoring noise correlations in optimal decoding. In practice, our study shows that the cost of ignoring noise correlations for studying brain computations and information integration, when evaluated using the communication notion of information, can be much lower than currently thought, thereby potentially saving experimental and computational resources, and contributing to develop simpler and more efficient neuroprosthetics and technological applications.

\section*{Acknowledgments} 

This work was supported by the Academy of Finland, Centre of Excellence in Inverse Problems (project number 213476) and Computational Sciences Program (project number 135198).

\bibliographystyle{harvard}

\begin{thebibliography}{64}
\expandafter\ifx\csname natexlab\endcsname\relax\def\natexlab#1{#1}\fi
\expandafter\ifx\csname url\endcsname\relax
  \def\url#1{\texttt{#1}}\fi
\expandafter\ifx\csname urlprefix\endcsname\relax\def\urlprefix{URL }\fi

\bibitem[{Abbott and Dayan(1999)}]{abbott1999}
Abbott, L., Dayan, P., 1999. The effect of correlated variability on the
  accuracy of a population code. Neural Comput. 11~(1), 91--101.

\bibitem[{Aflalo et~al.(2015)Aflalo, Kellis, Klaes, Lee, Shi, Pejsa, Shanfield,
  {Hayes-Jackson}, Aisen, Heck, Liu, and Andersen}]{aflalo2015}
Aflalo, T., Kellis, S., Klaes, C., Lee, B., Shi, Y., Pejsa, K., Shanfield, K.,
  {Hayes-Jackson}, S., Aisen, M., Heck, C., Liu, C., Andersen, R., 2015.
  Decoding motor imagery from the posterior parietal cortex of a tetraplegic
  human. Science 348~(6237), 906--910.

\bibitem[{Akam and Kullmann(2014)}]{akam2014}
Akam, T., Kullmann, D., 2014. Oscillatory multiplexing of population codes for
  selective communication in the mammalian brain. Nat. Rev. Neurosci. 15~(2),
  111--122.

\bibitem[{Averbeck et~al.(2006)Averbeck, Latham, and Pouget}]{averbeck2006}
Averbeck, B., Latham, P., Pouget, A., 2006. Neural correlations, population
  coding and computation. Nat. Rev. Neurosci. 7~(5), 358--366.

\bibitem[{Averbeck and Lee(2006)}]{averbeck2006b}
Averbeck, B., Lee, D., 2006. Effects of noise correlations on information
  encoding and decoding. J Neurophysiol 95~(6), 3633--3644.

\bibitem[{Babiloni and Astolfi(2014)}]{babiloni2014}
Babiloni, F., Astolfi, L., 2014. Social neuroscience and hyperscanning
  techniques: past, present and future. Neurosci Biobehav Rev 44, 76--93.

\bibitem[{Bialek(1987)}]{bialek1987}
Bialek, W., 1987. Physical limits to sensation and perception. Annu. Rev.
  Biophys. Biophys. Chem. 16~(1), 455--478.

\bibitem[{Bialek et~al.(1991)Bialek, Rieke, {de Ruyter van Steveninck}, and
  Warland}]{bialek1991}
Bialek, W., Rieke, F., {de Ruyter van Steveninck}, R., Warland, D., 1991.
  Reading a neural code. Science 252~(5014), 1854--1857.

\bibitem[{Bouton et~al.(2016)Bouton, Shaikhouni, Annetta, Bockbrader,
  Friedenberg, Nielson, Sharma, Sederberg, Glenn, Mysiw, Morgan, Deogaonkar,
  and Rezai}]{bouton2016}
Bouton, C., Shaikhouni, A., Annetta, N., Bockbrader, M., Friedenberg, D.,
  Nielson, D., Sharma, G., Sederberg, P., Glenn, B., Mysiw, W., Morgan, A.,
  Deogaonkar, M., Rezai, A., 2016. Restoring cortical control of functional
  movement in a human with quadriplegia. Nature.

\bibitem[{Brenner et~al.(2000)Brenner, Strong, Koberle, Bialek, and {de Ruyter
  van Steveninck}}]{brenner2000}
Brenner, N., Strong, S., Koberle, R., Bialek, W., {de Ruyter van Steveninck},
  R., 2000. Synergy in a neural code. Neural Comput. 12~(7), 1531--1552.

\bibitem[{Casella and Berger(2002)}]{casella2002}
Casella, G., Berger, R., 2002. Statistical Inference, second edition Edition.
  Duxbury.

\bibitem[{Cohen and Kohn(2011)}]{cohen2011}
Cohen, M.~R., Kohn, A., 2011. Measuring and interpreting neuronal correlations.
  Nat Neurosci 14~(7), 811--819.

\bibitem[{Cover and Thomas(2006)}]{cover2006}
Cover, T., Thomas, J., 2006. Elements of Information Theory, second edition
  Edition. Wiley-interscience.

\bibitem[{Delis et~al.(2013)Delis, Berret, Pozzo, and Panzeri}]{delis2013}
Delis, I., Berret, B., Pozzo, T., Panzeri, S., 2013. A methodology for
  assessing the effect of correlations among muscle synergy activations on
  task-discriminating information. Front Comput Neurosci 7, 54.

\bibitem[{Duda et~al.(2000)Duda, Hart, and Stork}]{duda2000}
Duda, R., Hart, P., Stork, D., 2000. Pattern Classification, second edition
  Edition. Wiley John \& Sons.

\bibitem[{Ernst and Banks(2002)}]{ernst2002}
Ernst, M., Banks, M., 2002. Humans integrate visual and haptic information in a
  statistically optimal fashion. Nature 415~(6870), 429--433.

\bibitem[{Eyherabide et~al.(2008)Eyherabide, Rokem, Herz, and
  Samengo}]{eyherabide2008}
Eyherabide, H., Rokem, A., Herz, A., Samengo, I., 2008. Burst firing is a
  neural code in an insect auditory system. Front. Comput. Neurosci. 2~(3).

\bibitem[{Eyherabide and Samengo(2010)}]{eyherabide2010FCN}
Eyherabide, H., Samengo, I., 2010. Time and category information in
  pattern-based codes. Front. Comput. Neurosci. 4, 145.

\bibitem[{Eyherabide and Samengo(2013)}]{eyherabide2013}
Eyherabide, H., Samengo, I., 2013. When and why noise correlations are
  important in neural decoding. J. Neurosci. 33~(45), 17921--17936.

\bibitem[{Eyherabide(2016)}]{eyherabide2016b}
Eyherabide, H.~G., 2016. Neural stochastic codes, encoding and decoding.
  Preprint. Available from: arxiv:1611.05080. Cited 6 January 2017.

\bibitem[{Fano(1961)}]{fano1961}
Fano, R., 1961. Transmission of Information. The M.I.T. Press.

\bibitem[{Gallager(1968)}]{gallager1968}
Gallager, R., 1968. Information theory and reliable communication. John Wiley
  and Sons, Inc., New York, USA.

\bibitem[{Gawne and Richmond(1993)}]{gawne1993}
Gawne, T., Richmond, B., 1993. How independent are the messages carried by
  adjacent inferior temporal cortical neurons? J. Neurosci. 13~(7), 2758--2771.

\bibitem[{Geisler(2011)}]{geisler2011}
Geisler, W., 2011. Contributions of ideal observer theory to vision research.
  Vision Res. 51~(7), 771--781.

\bibitem[{Gross et~al.(2013)Gross, Hoogenboom, Thut, Schyns, Panzeri, Belin,
  and Garrod}]{gross2013}
Gross, J., Hoogenboom, N., Thut, G., Schyns, P., Panzeri, S., Belin, P.,
  Garrod, S., 2013. Speech rhythms and multiplexed oscillatory sensory coding
  in the human brain. PLoS Biol. 11~(12), e1001752.

\bibitem[{Hari and Kujala(2009)}]{hari2009}
Hari, R., Kujala, M., 2009. Brain basis of human social interaction: from
  concepts to brain imaging. Physiol Rev 89, 453--479.

\bibitem[{Harvey et~al.(2013)Harvey, Saal, {Dammann III}, and
  Bensmaia}]{harvey2013}
Harvey, M., Saal, H., {Dammann III}, J., Bensmaia, S., 2013. Multiplexing
  stimulus information through rate and temporal codes in primate somatosensory
  cortex. PLoS Biol. 11~(5), e1001558.

\bibitem[{Huk(2012)}]{huk2012}
Huk, A., 2012. Multiplexing in the primate motion pathway. Vision Res. 62,
  173--180.

\bibitem[{Ince et~al.(2010)Ince, Senatore, Arabzadeh, Montani, Diamond, and
  Panzeri}]{ince2010}
Ince, R., Senatore, R., Arabzadeh, E., Montani, F., Diamond, M., Panzeri, S.,
  2010. Information-theoretic methods for studying population codes. Neural
  Netw. 23~(6), 713--727.

\bibitem[{Jaynes(2003)}]{jaynes2003}
Jaynes, E., 2003. Probability Theory: the Logic of Science. Cambridge
  university press.

\bibitem[{Knill and Richards(1996)}]{knill1996}
Knill, D., Richards, W., 1996. Perception as Bayesian Inference. Cambridge
  University Press.

\bibitem[{Landy et~al.(1995)Landy, Maloney, Johnston, and Young}]{landy1995}
Landy, M., Maloney, L., Johnston, E., Young, M., 1995. Measurement and modeling
  of depth cue combination: in defense of weak fusion. Vision Res. 35~(3),
  389--412.

\bibitem[{Latham and Nirenberg(2005)}]{latham2005}
Latham, P., Nirenberg, S., 2005. Synergy, redundancy, and independence in
  population codes, revisited. J. Neurosci. 25~(21), 5195--5206.

\bibitem[{Latham and Roudi(2013)}]{latham2013}
Latham, P., Roudi, Y., 2013. Role of correlations in population coding. CRC
  Press, Ch.~7, pp. 121--138.

\bibitem[{Meister and Hosoya(2001)}]{meister2001}
Meister, M., Hosoya, T., 2001. Are retinal ganglion cells independent encoders?
  Preprint.
\newline\urlprefix\url{http://www.gatsby.ucl.ac.uk/\~pel/critics/Meister\_2001\_Encoders.pdf}

\bibitem[{Meister et~al.(1995)Meister, Lagnado, and Baylor}]{meister1995}
Meister, M., Lagnado, L., Baylor, D., 1995. Concerted signaling by retinal
  ganglion cells. Science 270~(5239), 1207--1210.

\bibitem[{Merhav et~al.(1994)Merhav, Kaplan, Lapidoth, and {Shamai
  Shitz}}]{merhav1994}
Merhav, N., Kaplan, G., Lapidoth, A., {Shamai Shitz}, S., 1994. On information
  rates for mismatched decoders 40~(6), 1953--1967.

\bibitem[{Meytlis et~al.(2012)Meytlis, Nichols, and Nirenberg}]{meytlis2012}
Meytlis, M., Nichols, Z., Nirenberg, S., 2012. Determining the role of
  correlated firing in large populations of neurons using white noise and
  natural scene stimuli. Vision Res. 70, 44--53.

\bibitem[{Montani et~al.(2007)Montani, Kohn, Smith, and Schultz}]{montani2007}
Montani, F., Kohn, A., Smith, M., Schultz, S., 2007. The role of correlations
  in direction and contrast coding in the primary visual cortex. J Neurosci
  27~(9), 2338--2348.

\bibitem[{Nirenberg et~al.(2001{\natexlab{a}})Nirenberg, Carcieri, Jacobs, and
  Latham}]{nirenberg2001}
Nirenberg, S., Carcieri, S., Jacobs, A., Latham, P., 2001{\natexlab{a}}.
  Retinal ganglion cells act largely as independent encoders. Nature
  411~(6838), 698--701.

\bibitem[{Nirenberg et~al.(2001{\natexlab{b}})Nirenberg, Carcieri, Jacobs, and
  Latham}]{nirenberg2001addmeister}
Nirenberg, S., Carcieri, S., Jacobs, A., Latham, P., 2001{\natexlab{b}}.
  Supplementary information.
\newline\urlprefix\url{physiology.med.cornell.edu/faculty/nirenberg/lab/pel/critics/Meister_SI.ps}

\bibitem[{Nirenberg and Latham(1998)}]{nirenberg1998}
Nirenberg, S., Latham, P., 1998. Population coding in the retina. Curr. Opin.
  Neurobiol. 8~(4), 488--493.

\bibitem[{Nirenberg and Latham(2003)}]{nirenberg2003}
Nirenberg, S., Latham, P., 2003. Decoding neuronal spike trains: how important
  are correlations? Proc. Natl. Acad. Sci. U.S.A. 100~(12), 7348--7353.

\bibitem[{Oizumi et~al.(2016)Oizumi, Amari, Yanagawa, Fujii, and
  Tsuchiya}]{oizumi2016}
Oizumi, M., Amari, S., Yanagawa, T., Fujii, N., Tsuchiya, N., 2016. Measuring
  integrated information from the decoding perspective. PLoS Comput. Biol.
  12~(1), e1004654.

\bibitem[{Oizumi et~al.(2009)Oizumi, Ishii, Ishibashi, Hosoya, and
  Okada}]{oizumi2009}
Oizumi, M., Ishii, T., Ishibashi, K., Hosoya, T., Okada, M., 2009. A general
  framework for investigating how far the decoding process in the brain can be
  simplified. Adv. Neural Inf. Process. Syst., 1225--1232.

\bibitem[{Oizumi et~al.(2010)Oizumi, Ishii, Ishibashi, Hosoya, and
  Okada}]{oizumi2010}
Oizumi, M., Ishii, T., Ishibashi, K., Hosoya, T., Okada, M., 2010. Mismatched
  decoding in the brain. J. Neurosci. 30~(13), 4815--4826.

\bibitem[{Oppenheim et~al.(1997)Oppenheim, Willsky, and Nawab}]{oppenheim1997}
Oppenheim, A., Willsky, A., Nawab, S., 1997. Signals and systems, second
  edition Edition. Prentice Hall.

\bibitem[{Oram et~al.(1998)Oram, F{\"o}ldi{\'a}k, Perrett, and
  Sengpiel}]{oram1998}
Oram, M., F{\"o}ldi{\'a}k, P., Perrett, D., Sengpiel, F., 1998. The 'ideal
  homunculus': decoding neural population signals. Trends Neurosci. 21~(6),
  259--265.

\bibitem[{Panzeri et~al.(2010)Panzeri, Brunel, Logothetis, and
  Kayser}]{panzeri2010}
Panzeri, S., Brunel, N., Logothetis, N., Kayser, C., 2010. Sensory neural codes
  using multiplexed temporal scales. Trends Neurosci. 33~(3), 111--120.

\bibitem[{Panzeri et~al.(2001)Panzeri, Golledge, Zheng, Tov{\'e}e, and
  Young}]{panzeri2001}
Panzeri, S., Golledge, H., Zheng, F., Tov{\'e}e, M., Young, M., 2001. Objective
  assessment of the functional role of spike train correlations using
  information measures. Vis. Cogn. 8~(3-5), 531--547.

\bibitem[{Pereda et~al.(2005)Pereda, Quiroga, and Bhattacharya}]{pereda2005}
Pereda, E., Quiroga, R., Bhattacharya, J., 2005. Nonlinear multivariate
  analysis of neurophysiological signals. Prog. Neurobiol. 77~(1--2), 1--37.

\bibitem[{Quiroga and Panzeri(2009)}]{quiroga2009}
Quiroga, R., Panzeri, S., 2009. Extracting information from neuronal
  populations: information theory and decoding approaches. Nat. Rev. Neurosci.
  10~(3), 173--185.

\bibitem[{Rolls and Treves(2011)}]{rolls2011}
Rolls, E., Treves, A., 2011. The neuronal encoding of information in the brain.
  Prog. Neurobiol. 95~(3), 448--490.

\bibitem[{Samengo(2002)}]{samengo2002}
Samengo, I., 2002. Estimating probabilities from experimental frequencies. Phys
  Rev E Stat Nonlin Soft Matter Phys 65~(4), 046124.

\bibitem[{Samengo and Treves(2000)}]{samengo2000}
Samengo, I., Treves, A., 2000. Representational capacity of a set of
  independent neurons. Phys. Rev. E 63~(1), 011910.

\bibitem[{Schneidman et~al.(2003)Schneidman, Bialek, and
  Berry}]{schneidman2003}
Schneidman, E., Bialek, W., Berry, M., 2003. Synergy, redundancy, and
  independence in population codes. J. Neurosci. 23~(37), 11539--11553.

\bibitem[{Schneidman et~al.(2011)Schneidman, Puchalla, Segev, Harris, Bialek,
  and Berry}]{schneidman2011}
Schneidman, E., Puchalla, J., Segev, R., Harris, R., Bialek, W., Berry, M.,
  2011. Synergy from silence in a combinatorial neural code. J. Neurosci.
  31~(44), 15732--15741.

\bibitem[{Shannon and Weaver(1949)}]{shannon1949}
Shannon, C., Weaver, W., 1949. The Mathematical Theory of Communication.
  University of Illinois Press.

\bibitem[{Simoncelli(2009)}]{simoncelli2009}
Simoncelli, E., 2009. Optimal estimation in sensory systems. The MIT Press, pp.
  525--535.

\bibitem[{Thomson and Kristan(2005)}]{thomson2005}
Thomson, E., Kristan, W., 2005. Quantifying stimulus discriminability: a
  comparison of information theory and ideal observer analysis. Neural
  Computation 17~(4), 741--778.

\bibitem[{Warland et~al.(1997)Warland, Reinagel, and Meister}]{warland1997}
Warland, D., Reinagel, P., Meister, M., 1997. Decoding visual information from
  a population of retinal ganglion cells. J. Neurophysiol. 78~(5), 2336--2350.

\bibitem[{Womelsdorf et~al.(2012)Womelsdorf, Lima, Vinck, Oostenveld, Singer,
  Neuenschwander, and Fries}]{womelsdorf2012}
Womelsdorf, T., Lima, B., Vinck, M., Oostenveld, R., Singer, W.,
  Neuenschwander, S., Fries, P., 2012. Orientation selectivity and noise
  correlation in awake monkey area v1 are modulated by the gamma cycle. Proc
  Natl Acad Sci U S A 109~(11), 4302--4307.

\bibitem[{Woodward and Davies(1952)}]{woodward1952}
Woodward, P., Davies, I., 1952. Information theory and inverse probability in
  telecommunication. Proc. I. E. E. 99~(58), 37--44.

\bibitem[{Zhang et~al.(2016)Zhang, Chen, Rasch, and Wu}]{zhang2016}
Zhang, W., Chen, A., Rasch, M., Wu, S., 2016. Decentralized multisensory
  information integration in neural systems. J. Neurosci. 36~(2), 532--547.

\end{thebibliography}

\end{document}